\documentclass[twocolumn]{custom_BJ}
\usepackage[english]{babel}
\usepackage[utf8]{inputenc}
\usepackage[T1]{fontenc}

\usepackage{graphicx}
\usepackage{combelow}
\usepackage{mathtools}

\usepackage{sidecap}

\usepackage[colorlinks,allcolors=black]{hyperref}

\usepackage{pdfpages}

\title{\vspace{-1cm}\centering\bfseries\Large Quantifying cell cycle regulation by tissue crowding}

\author{\centering \normalsize Carles Falc\'o$^{1,}$\footnote{falcoigandia@maths.ox.ac.uk}\;, Daniel J. Cohen$^{2,3}$, Jos\'e A. Carrillo$^{1}$, Ruth E. Baker$^{1}$}
\date{}

\begin{document}

\twocolumn[\maketitle\par\vspace{-1.3cm}
\begin{center}
{$^{1}$Mathematical Institute, University of Oxford, OX2 6GG Oxford, United Kingdom}

{$^2$Department of Mechanical and Aerospace Engineering, Princeton University, Princeton, NJ, 08544, USA}

{$^3$Department of Chemical and Biological Engineering, Princeton University, Princeton, NJ, 08544, USA}

{$^*$Correspondence: falcoigandia@maths.ox.ac.uk}
\end{center}

The spatiotemporal coordination and regulation of cell proliferation is fundamental in many aspects of development and tissue maintenance. Cells have the ability to adapt their division rates in response to mechanical constraints, yet we do not fully understand how cell proliferation regulation impacts cell migration phenomena. Here, we present a minimal continuum model of cell migration with cell cycle dynamics, which includes density-dependent effects and hence can account for cell proliferation regulation. By combining minimal mathematical modelling, Bayesian inference, and recent experimental data, we quantify the impact of tissue crowding across different cell cycle stages in epithelial tissue expansion experiments. Our model suggests that cells sense local density and adapt cell cycle progression in response, during G1 and the combined S/G2/M phases, providing an explicit relationship between each cell cycle stage duration and local tissue density, which is consistent with several experimental observations. Finally, we compare our mathematical model predictions to different experiments studying cell cycle regulation and present a quantitative analysis on the impact of density-dependent regulation on cell migration patterns. Our work presents a systematic approach for investigating and analysing cell cycle data, providing mechanistic insights into how individual cells regulate proliferation, based on population-based experimental measurements.
\par\vspace{2ex}]

\section*{Introduction}
The coordination of cell proliferation across space and time is crucial for the emergence of collective cell migration, which plays a fundamental role in development, including tissue formation and morphogenesis, and also at later stages for tissue regeneration and homeostasis. Cells adapt their division rates in response to mechanical constraints within tissues \cite{JORGENSEN2004R1014,streichan2014spatial}, allowing cell populations to self-organise and eventually form and maintain tissues and complex structures. Moreover, disruptions in the control of cell proliferation often result in tumour formation \cite{massague2004g1,MCCLATCHEY2012685,otto2017cell}. Although significant experimental efforts have been devoted to understand the mechanical regulation of cell proliferation  \cite{gupta2022mechanical} and its interplay with collective cell migration, existing mathematical models have failed to describe these constraints and how they affect cell cycle progression \cite{vittadello2018mathematical,simpson2020practical,gavagnin2019invasion}.

\begin{figure*}[hbt!]
    \centering
    \includegraphics[width = \textwidth]{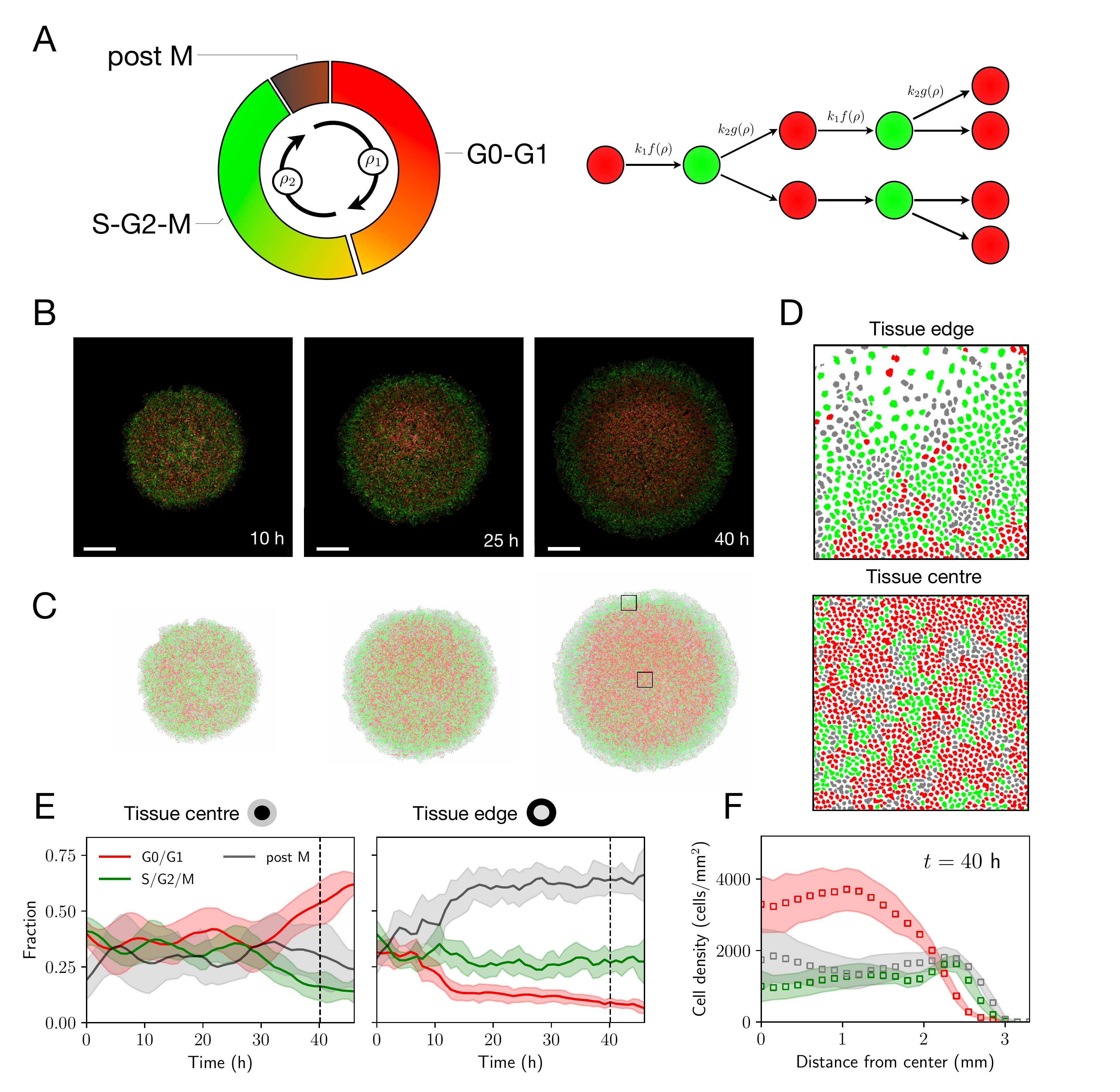}
    \caption{(A) Schematics of the FUCCI cell cycle marker system and model conceptualisation. Transitions in the model given by Eqs.~\eqref{eq: full model} are regulated by the \emph{crowding functions} $f(\rho)$ and $g(\rho)$, dependent on the total cell density $\rho = \rho_1 + \rho_2$.  (B) FUCCI fluorescence images from the experiments of Heinrich et al. \cite{heinrich2020size} at different time points (adapted). Initial tissue diameter $\sim 3.4$ mm. Scale bars correspond to 1 mm. (C) Segmented data showing G1 (red), S/G2/M (green), and post-mitotic (gray) cells. Note that in the model we combine post-mitotic cells and cells in G1. (D) Zoomed-in segmented data at the tissue edge and centre, corresponding to the black squares in (C). (E) Fraction of cell-cycle state cells in the tissue centre and in the tissue edge --- defined as regions extending $\sim200$ $\mu$m from the tissue center and tissue edge, respectively. (F) Density profiles in polar coordinates at $t=40$ h, showing cells in G1, S/G2/M, and post-mitotic cells. (E) and (F) show the average of eleven independent tissue expansions with the same experimental initial condition, with shaded regions indicating one standard deviation with respect to the mean.}
    \label{fig: figure 1}
\end{figure*}

In order to understand cell proliferation regulation, numerous experimental studies have explored how spatial and mechanical constraints within tissues affect different stages of the cell cycle. The cell cycle consists of four main stages, namely: the G1 phase, where cells grow and prepare for DNA replication; the S phase, during which DNA synthesis occurs; the G2 phase, characterised by further cell growth and preparation for mitosis; and finally, the M phase, where cell division takes place. Cells can also exit the cell cycle and enter G0, where they become quiescent. The experimental visualisation of cell cycle stages can be achieved via the widely used FUCCI cell-cycle marker \cite{sakaue2008visualizing}, which consists of red and green fluorescent proteins that are fused to proteins Cdt1 and Geminin, respectively. Cdt1 exhibits elevated levels during the G0/G1 phase and decreased levels throughout the remaining cell cycle stages, whereas Geminin shows high expression during the S, G2, and M phases; allowing thus to distinguish between these different stages --- see Fig.~\ref{fig: figure 1}. Several extensions of the FUCCI system exist now \cite{ridenour2012cycletrak}; for instance FUCCI4 allows for the simultaneous visualisation of the G1, S, G2, and M phases \cite{bajar2016fluorescent}.

Experimental studies of cell migration are often performed in epithelia due to their strong cell-cell adhesion which gives rise to collective and cohesive motion. Moreover, they play a fundamental role in multicellular organisms as they serve as protective layers for various body surfaces and organs. Epithelial cell proliferation is regulated by mechanical forces, which can accelerate, delay, arrest, or re-activate the cell cycle. In particular, extensive research has focused on the G1-S boundary, revealing that intercellular tension can favour this transition \cite{uroz2018regulation}, while tissue pressure can halt progression based on crowding \cite{streichan2014spatial}.

More generally, the extracellular regulation of switches from G0 and G1, and within substages of G1, has been well-known for many years \cite{pardee1989g1}. However, and contrary to initial assumptions, cells also have the ability to regulate progression through stages of the cell cycle following the G1-S transition in response to external cues. These external signals might involve not only mechanical forces \cite{godard2019cell}, but also nutrients and growth factors \cite{mckeown2019nutrient,CELORA2022111104}. In epithelia, this question was explored recently by Donker et al. \cite{donker2022mechanicalG2}, revealing a mechanical checkpoint in G2 which controls cell division. In particular, this checkpoint allows cells to regulate progression through G2, via sensing of local density, explaining why dense regions in epithelia contain groups of cells that are temporarily halted in G2.

Experimental studies employing FUCCI and variations of it have thus successfully linked mechanical constraints to cell cycle progression. These studies have employed qualitative analysis, direct measurements of cell cycle stage durations \cite{donker2022mechanicalG2}, or metrics associated with cell cycle progression, such as cell area \cite{streichan2014spatial}, and Geminin/Cdt1 or EdU  signals \cite{suh2023cadherin,hollring2023capturing}. However, these approaches omit a quantitative comparison between model and data, hence limiting the depth of mechanistic insights that can be derived.

Here, we present a quantitative investigation into the mechanical regulation of cell cycle progression by sensing of local tissue density First, we construct a mathematical model of cell cycle dynamics that accurately captures the impact of tissue crowding on cell cycle progression.
By combining minimal mathematical modelling, Bayesian inference, and recent experimental data \cite{heinrich2020size}, we provide further evidence, consistent with previous experimental studies \cite{streichan2014spatial,donker2022mechanicalG2}, that density-dependent effects operate throughout the cell cycle and together serve as a regulating mechanism for the growth of epithelial tissues. Our work thus constitutes a systematic approach towards the quantification of density-dependent effects regulating cell cycle progression. Moreover, the obtained parameter estimates reveal an explicit relation between the duration of different cell cycle stages and tissue density, which is consistent with the experimental measurements of Donker et al. \cite{donker2022mechanicalG2}.

\section*{Methods}

\subsection*{Mathematical models of cell cycle dynamics}

We build on the model proposed by Vittadello et al. \cite{vittadello2018mathematical} to describe two cell populations, $\rho_1(\mathbf{x},t)$ and $\rho_2(\mathbf{x},t)$, in different stages of the cell cycle. We represent by $\rho_1$ the density of cells that are in G0/G1, while $\rho_2$ gives the density of cells in the S/G2/M phases of the cell cycle --- see Fig.~\ref{fig: figure 1}. In the model, cell motility is described via linear diffusion, with a diffusion constant $D>0$ for both cell populations \cite{vittadello2020examining}. In order to effectively capture density-dependent effects controlling cell cycle progression, we assume that the transitions between different cell cycle stages are regulated by two \emph{crowding functions}, $f(\rho)$ and $g(\rho)$, which depend on the total cell density $\rho = \rho_1 + \rho_2$. In particular, the transition rate from G1 to S is given by $k_1f(\rho)$, while the division rate (from S/G2/M to G1) is given by $k_2g(\rho)$, where $k_1, k_2>0$ are intrinsic rates of cell cycle progression. With this, the model reads
\begin{equation}\label{eq: full model}
    \begin{split}
        \partial_t\rho_1 & = D\Delta\rho_1 - k_1\rho_1f(\rho) + 2k_2\rho_2g(\rho),
        \\
        \partial_t\rho_2 &= D\Delta\rho_2 + k_1\rho_1f(\rho) - k_2\rho_2g(\rho),
    \end{split}
\end{equation}
where the factor of two in the equation for $\rho_1$ represents cell division into two daughter cells, and $\Delta = \sum_{i=1}^d\partial^2_{x_{i}}$ is the Laplacian operator in dimension $d$. These equations are solved first in polar coordinates (assuming radial symmetry in two spatial dimensions, $d = 2$) to describe epithelial tissue expansion experiments, and then in one spatial dimension ($d=1$) to study travelling wave behaviour, and the impact of tissue crowding on cell migration phenomena.

In order to accurately capture density-dependent effects regulating cell cycle progression, we assume that $f$ and $g$ are non-increasing functions of the total density $\rho$. Again, this is motivated by the experimental observations of Streichan et al. \cite{streichan2014spatial} and Donker et al. \cite{donker2022mechanicalG2}. Furthermore, we assume $f(0) = g(0) = 1$, so that $k_1$ and $k_2$ represent density-independent transition rates. Note that setting $f = g\equiv 1$ gives rise to an exponential growth model (i.e. no dependence on density). On the other hand, choosing $f\equiv 1$ and $g(\rho) = (1 - \rho/K)_+$ we recover the Vittadello et al. model \cite{vittadello2018mathematical}. Here, we  assume that $f(\rho)$ and $g(\rho)$ decrease linearly with the total cell density so that
\begin{equation}\label{eq: crowding functions}
 f(\rho) = \left(1-\frac{\rho}{K_1}\right)_+,\quad g(\rho) = \left(1-\frac{\rho}{K_2}\right)_+\,,
\end{equation}
where $K_1, K_2>0$ are constants controlling the duration of G1 and the S/G2/M phases, respectively, and $(z)_+ = \max(z,0)$. The specific form of these \emph{crowding functions} is chosen here for simplicity, although other functions sharing the same properties show similar qualitative behaviour. 

We follow a Bayesian approach \cite{Hines2014DeterminationOP,simpson2020practical,falco2023quantifying,pypesto} to calibrate the model given in Eqs.~\eqref{eq: full model}. In particular, given experimental measurements of the cell densities $\{\rho_k^\mathcal{D}(\mathbf{x}_i,t_j)\}_{i,j}$ for $k = 1,2$, and a vector of model parameters $\theta = (D,k_1,k_2,K_1,K_2)$, we estimate the posterior probability distribution $p(\theta|\rho^\mathcal{D})$, which gives the probability density for the model parameters taking specific values. The posterior distribution, thus, can be used to quantify the uncertainty associated with specific parameter values, given the experimental observation. We refer the reader to the Supplementary Information for more details on Bayesian inference.

\section*{Results}

\subsection*{Tissue expansion experiments}

\begin{figure*}[hbt!]
    \centering
    \includegraphics[width = \textwidth]{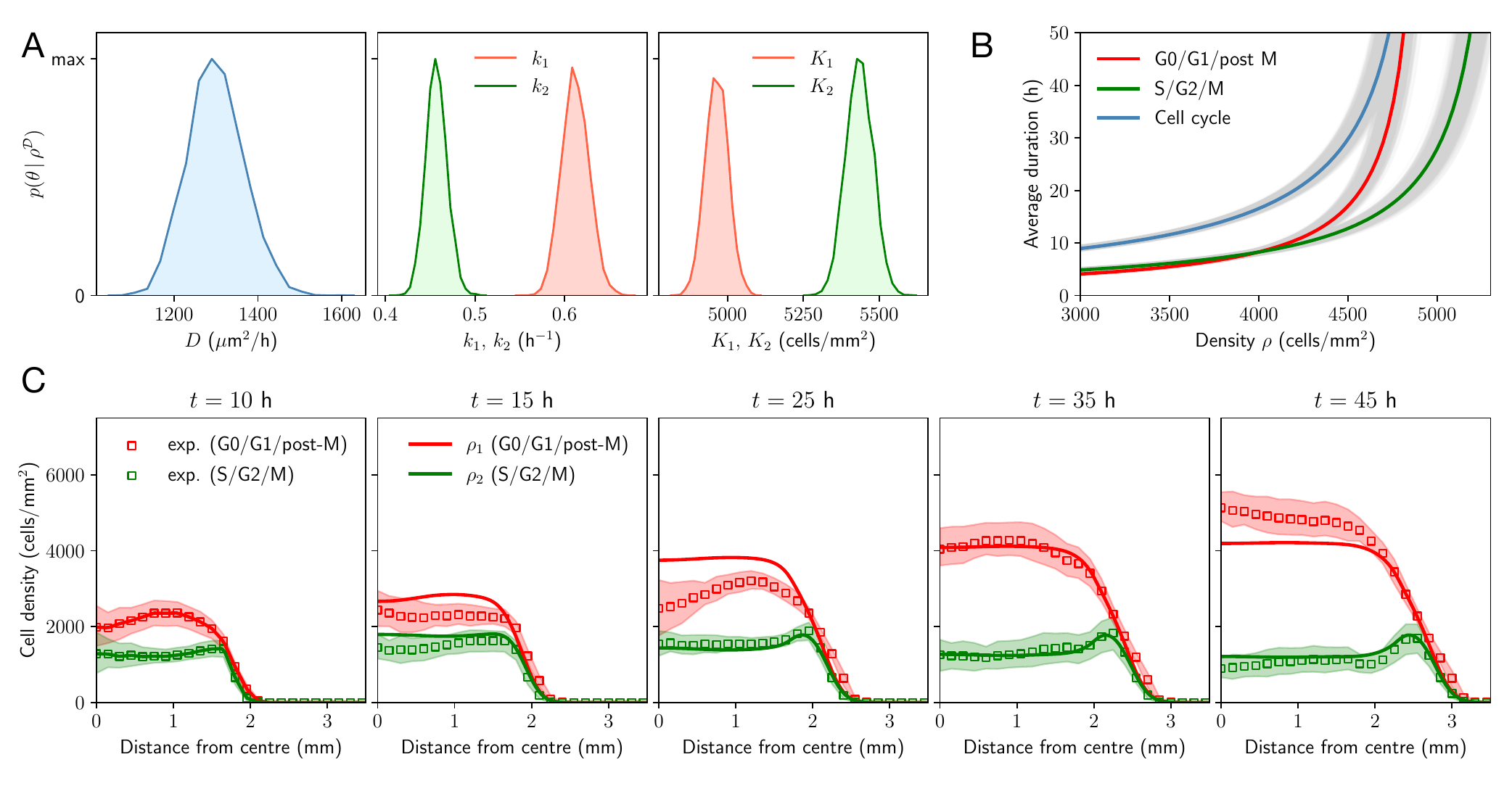}
    \caption{Density-dependent effects regulate cell cycle dynamics in epithelial tissue expansion experiments \cite{heinrich2020size}. Parameter estimation and model-data comparison for the model given by Eqs.~\eqref{eq: full model} and~\ref{eq: crowding functions}. (A) Univariate marginal posterior distributions for the model parameters. Posterior modes are given by $(D,k_1,k_2,K_1,K_2) = (1300\pm 66 $ $\mu$m$^2/$h$,0.612\pm0.015 $ h$^{-1}, 0.457\pm0.011 $ h$^{-1}, 4965\pm38 $ cells$/$mm$^2, 5435\pm45 $ cells$/$mm$^2)$, where errors correspond to one standard deviation. (B) Estimated duration of the G0/G1/post M (red) and S/G2/M (green) phases, as well as the whole cell cycle (blue), as a function of cell densities. Solid lines correspond to posterior modes and shaded regions are obtained sampling from the posterior distribution. (C) Comparing data and model predictions. Squares represent the estimated cell density obtained by averaging eleven experimental realisations, which we use to calibrate the model. Shaded regions denote one standard deviation with respect to the mean -- see Supplementary Fig. 2 for confidence intervals in the model predictions. Numerical simulations in polar coordinates were obtained by using the posterior modes as parameter values, and no-flux boundary conditions --- for details on the numerical scheme we refer to the Supplementary Information. In order to minimise the effects of the stencil removal on cell behaviour, the initial condition corresponds to the experimental density profile ten hours after stencil removal.}
    \label{fig: figure 2}
\end{figure*}

We compare our model predictions to the experiments performed by Heinrich et al. \cite{heinrich2020size} studying the expansion and growth dynamics of a single circular epithelial tissue --- see Fig.~\ref{fig: figure 1}B. In these experiments, MDCK cells expressing the FUCCI markers are cultured in a silicone stencil for 18 hours and, after stencil removal, the cell population is allowed to freely expand for 46 hours. Given that the average cell cycle duration for MDCK cells is around 16 hours, this enables each cell to potentially undergo 2-3 cell divisions during the experiment. Local densities are then quantified by segmenting the fluorescence images in ImageJ and counting the number of nucleus centroids --- Fig.~\ref{fig: figure 1}C. Note that post-mitotic cells do not fluoresce and appear dark, which makes the FUCCI system unreliable for cell counting. To quantify the density of post-mitotic cells, Heinrich et al. used a convolutional neural network to identify nuclei from phase contrast images \cite{lachance2020practical} --- see \cite{heinrich2020size} for more details.  Moreover, and in line with previous work \cite{falco2023quantifying}, the model takes as initial condition the quantified density profile ten hours after stencil removal, so that the impact of the stencil on the dynamics is reduced. Note that after this time, cell densities near the tissue centre are relatively high ($\sim 3500$ cells/mm$^2$, which corresponds to around 50-70$\%$ of the maximum saturation density for MDCK \cite{heinrich2020size, falco2023quantifying}) and a fraction of cells in this region are likely to be found in a quiescent state due to contact inhibition of locomotion and proliferation  \cite{puliafito2012collective}.

The experiments by Heinrich et al. \cite{heinrich2020size} reveal a higher density of cells in G0/G1 at the centre of the tissue, where the total cell density is also higher --- see Fig.~\ref{fig: figure 1}D-E. The tissue edge, in contrast, is characterised by a larger number of cells which are preparing to divide (green) or are directly post-mitotic (gray). This agrees with previous observations of epithelial cells, which are known to control progression from G1 to S in response to spatial constraints \cite{streichan2014spatial}. Note, however, that the density of cells in S/G2/M in the tissue centre is low but non-zero, even at later times in the experiment --- see Fig.~\ref{fig: figure 1}D-F --- as observed also by Donker et al. \cite{donker2022mechanicalG2}.

For the sake of simplicity, here we consider post-mitotic cells (gray in Fig.~\ref{fig: figure 1}) and cells in G0/G1, as one single cell population.  Quantifying post-mitotic cell density is crucial in order to estimate both $K_1$ and $K_2$ in Eqs.~\eqref{eq: crowding functions}, given that these parameters are measures of contact inhibition of proliferation, typically associated with regions of higher cell density \cite{warne2017optimal}.

To calibrate the model, we fit to the estimated cell density obtained by averaging eleven experimental realisations. We show the univariate marginal posterior distributions corresponding to the model parameters in 
Fig.~\ref{fig: figure 2}A, confirming that all model parameters are practically identifiable. In particular, all marginal posteriors show well-defined and unimodal distributions, with a relatively narrow variance. For more details on model calibration we refer to the Supplementary Information (Supplementary Fig. 1).

The posterior distributions in Fig.~\ref{fig: figure 2}A are not only useful to inform further model predictions, but also give insights into the fundamental mechanisms underlying cell proliferation. In particular, given the intrinsic transition rate from G1 to S, $k_1$, and the constant $K_1$ in Eqs.~\eqref{eq: crowding functions}, we can estimate the average duration of the combined G1/post-M phase, for a given fixed density $\rho$, as $ 1/k_1f(\rho) = 1/(k_1(1-\rho/K_1)_+)$. Analogously, the estimated average duration of the S/G2/M phases is given by $1/k_2g(\rho)= 1/(k_2(1-\rho/K_2)_+)$. Note, however, that these are only estimates of the timescales associated with different cell cycle stages. Put together, these estimates predict for a range of densities between 4000-4500 cells$/$mm$^2$, a population doubling time of 14-20 hours. In Fig.~\ref{fig: figure 2}B we plot these timescales as a function of the density $\rho$, observing how the duration of the different cell cycle stages increases with density. These results confirm again, in line with previous experimental measurements \cite{streichan2014spatial,donker2022mechanicalG2}, that cell cycle dynamics are tightly regulated by density-dependent effects. In particular, these estimates are consistent with the experimental measurements of Donker et al. \cite{donker2022mechanicalG2} -- taking into account that the initial cell densities in our datasets are around $\rho\sim 3500$ cells/mm$^2$. At very low densities, however, our estimates predict a relatively short cell cycle duration. This suggests that the shape of the \emph{crowding functions} $f$ and $g$ might be closer to a constant function in this regime.

In Fig.~\ref{fig: figure 2}C we show numerical solutions of the model (Eqs.~\eqref{eq: full model} and~\eqref{eq: crowding functions}), taking the posterior modes as parameter values. These confirm that the model can describe cell cycle dynamics inside expanding epithelial tissues. Notably, the model captures the tissue expansion speed, as well as the S/G2/M density peak near the edge of the tissue, which results from density-dependent effects regulating the cell cycle. We also note that this type of density profile is possible in the model when crowding-dependent effects are stronger in the early stages of the cell cycle (G1/post M) and weaker in the latter ones (S/G2/M). In terms of Eqs.~\eqref{eq: crowding functions} this requires having $K_1<K_2$, which is correctly identified from the data.

We observe that the model overestimates the experimental density for early times of the experiment and, as a result of the model fit, underestimates it at later times. This is likely due to the transient behaviour that cells exhibit immediately after stencil removal \cite{JIN2016136,jin2017logistic}, which could have an impact on cell behaviour even after the first ten hours of expansion, as suggested also in previous studies \cite{falco2023quantifying}. However, we emphasise that tissue edge motion can be well described by the model. 

A similar behaviour is reported when the model is compared to a second set of experiments performed by Heinrich et al. \cite{heinrich2020size}. In this case, we use the obtained parameter estimates to describe the expansion of initially smaller epithelial monolayers (initial diameter $\sim 1.7$ mm). We highlight that the mathematical model can capture the expansion dynamics near the tissue edge as well as the expansion speed (see Supplementary Fig. 3), even though model parameters were inferred from the large tissue expansions.

\subsection*{Tissue colonisation experiments}

Our model, together with the experiments of Heinrich et al. \cite{heinrich2020size}, reveals the intrinsic connection between tissue crowding and cell cycle progression, showcasing how this interplay can give rise to spatiotemporal patterns of cell proliferation in growing tissues. Next, we show how the model can be used to study and describe similar patterns observed in several other experimental studies using FUCCI and variants of it.

\begin{figure*}[hbt!]
    \centering
    \includegraphics[width = .82\textwidth]{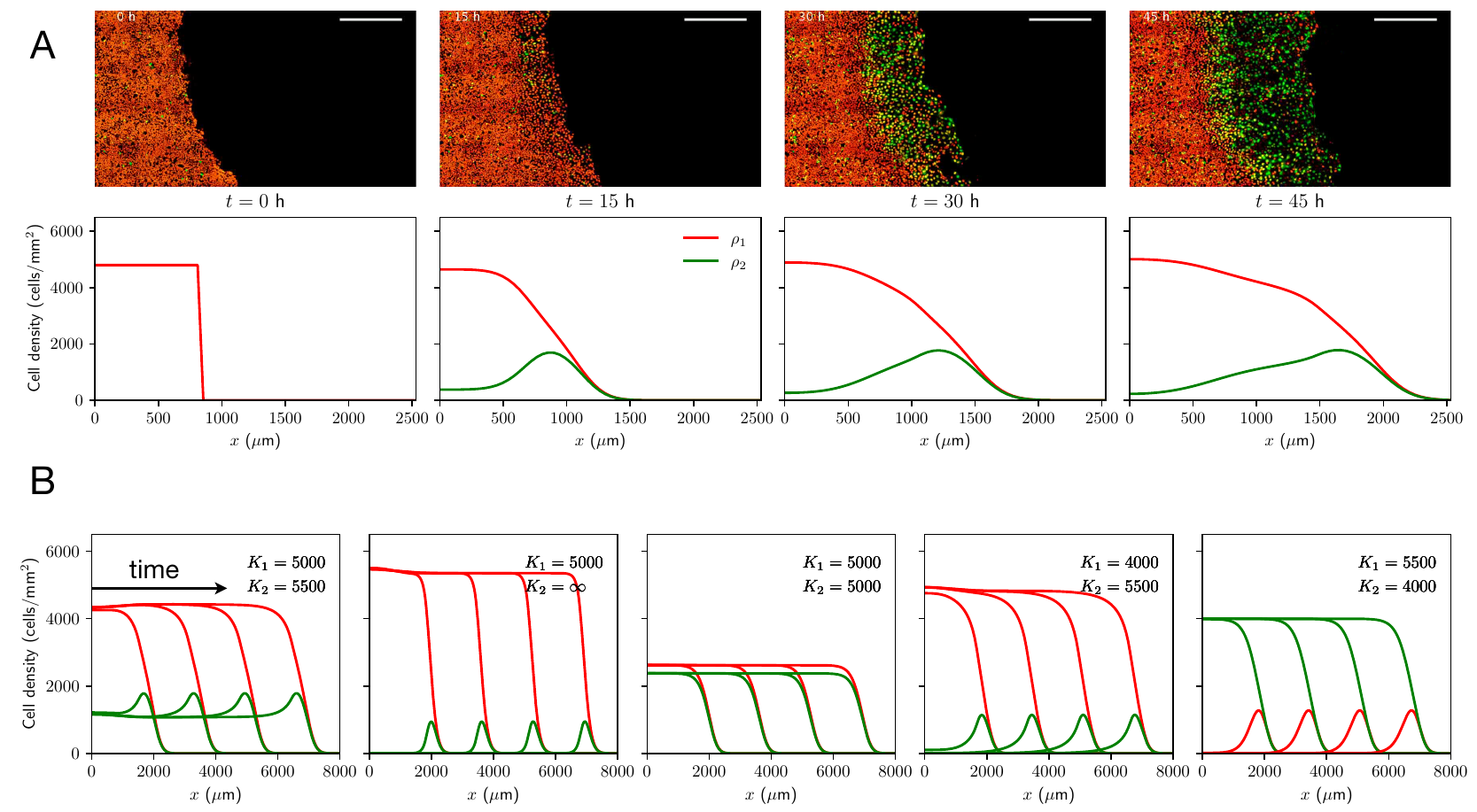}
    \caption{Cell cycle regulation by tissue crowding impacts cell migration. (A) Comparison with the tissue colonisation experiments of Streichan et al. \cite{streichan2014spatial} (top row, adapted with permission). Scale bars correspond to 500 $\mu$m. 
Bottom row shows numerical solutions of Eqs.~\eqref{eq: full model} on a one-dimensional domain of length 3000 $\mu$m with no-flux boundary conditions, and initial conditions: $\rho_1(x,0) =  4800$ cells/mm$^2$ for $x<850$ $\mu$m and $\rho_1(x,0) = 0$  cells/mm$^2$ otherwise; $\rho_2(x,0)= 0$. Parameter values correspond to the posterior modes in Fig.~\ref{fig: figure 2}. (B) Travelling wave solutions of  Eqs.~\eqref{eq: full model} for different values of $K_1$ and $K_2$ and at time points $t = 50,\, 100,\, 150,\,200$ h. Units of $K_1$ and $K_2$ are cells$/$mm$^2$. Initial conditions: $\rho_1(x,0) = \rho_2(x,0) = 500$ cells/mm$^2$ for $x<850$ $\mu$m, and  $\rho_1(x,0) = \rho_2(x,0) = 0$ cells/mm$^2$ otherwise. In all cases, all parameters except for $K_1$ and $K_2$ are fixed (taken from posterior modes).}
    \label{fig: figure 3}
\end{figure*}

Streichan et al. \cite{streichan2014spatial}  show, using a tissue barrier assay, how  the cell cycle can be reactivated by allowing cells to migrate and colonise free space -- see top row in Fig.~\ref{fig: figure 3}A. These experiments are initialised by growing MDCK-2 FUCCI cells in the G0/G1 phase within a removable barrier. After barrier removal, the tissue quickly colonises the available space, and cells behind the barrier, which were initially in G0/G1, reactivate their cycle by entering S phase. On the other hand, cells located further behind the barrier remain at high density and do not progress through the cell cycle. 

By solving numerically Eqs.~\eqref{eq: full model} on a one-dimensional domain -- see bottom row in Fig.~\ref{fig: figure 3}A -- we immediately observe how a model accounting for density-dependent regulation predicts similar behaviour to that observed experimentally\footnote{Note that the experimental images from Streichan et al. \cite{streichan2014spatial} do not show post-mitotic cells which appear dark in the FUCCI system, and that the total cell density (including post-mitotic cells) was used to estimate the parameters in the model.}. In particular, and as inferred from the experimental data of Heinrich et al. \cite{heinrich2020size}, the calibrated model predicts that crowding-dependent effects have a greater impact at the G1-S transition, compared to the S/G2/M phases. In terms of the model and the choice of \emph{crowding functions} (Eqs.~\eqref{eq: crowding functions}), this once again requires $K_1< K_2$.

\subsection*{Cell cycle regulation and cell migration}

Given that assuming $K_1<K_2$ seems necessary in order to obtain biologically realistic model predictions, 
what role does tissue crowding play in shaping cell migration patterns? We explore this question by varying the values of $K_1$ and $K_2$ in Eqs.~\eqref{eq: crowding functions} -- Fig.~\ref{fig: figure 3}B.
First, we observe that the density of S/G2/M peaks near the tissue edge when $K_2>K_1>0$ and remains low in the tissue bulk as long as $K_2\gg K_1$. However, for $K_2\sim K_1$, the height of this peak decreases and the fraction of S/G2/M cells in the tissue bulk increases.
On the other hand, when we assume a higher influence of density during S/G2/M relative to G1/post-M ($K_2 < K_1$), we observe that the tissue centre shows a higher fraction of cells in S/G2/M, in contrast with previously reported observation of contact inhibition of proliferation \cite{puliafito2012collective}. 

The numerical solutions in Fig.~\ref{fig: figure 3}B suggest that low-density initial conditions lead to travelling wave solutions in one spatial dimension: $\rho_1(x-ct)$, $\rho_2(x-ct)$, with $c>0$ being the wave speed, and $x$ denoting the spatial coordinate. Standard arguments -- see Supplementary Information --  predict the existence of a minimum travelling wave speed in terms of only three model parameters
\begin{equation}
     c_\text{min} = \sqrt{2D\left(-k_1-k_2+\sqrt{k_1^2+k_2^2+6k_1k_2}\,\right)}\,.\label{eq: minimum travelling wave speed}
\end{equation}
Interestingly, this suggests that the invasion speed is independent of cell cycle regulation, and only depends on cell motility ($D$), and the intrinsic, density-independent growth rates ($k_1$ and $k_2$). 
However, we highlight that, as shown in the figure, crowding constraints play an important role in shaping collective migration patterns. 

\begin{figure*}[hbt!]
    \centering
    \includegraphics[width = .9\textwidth]{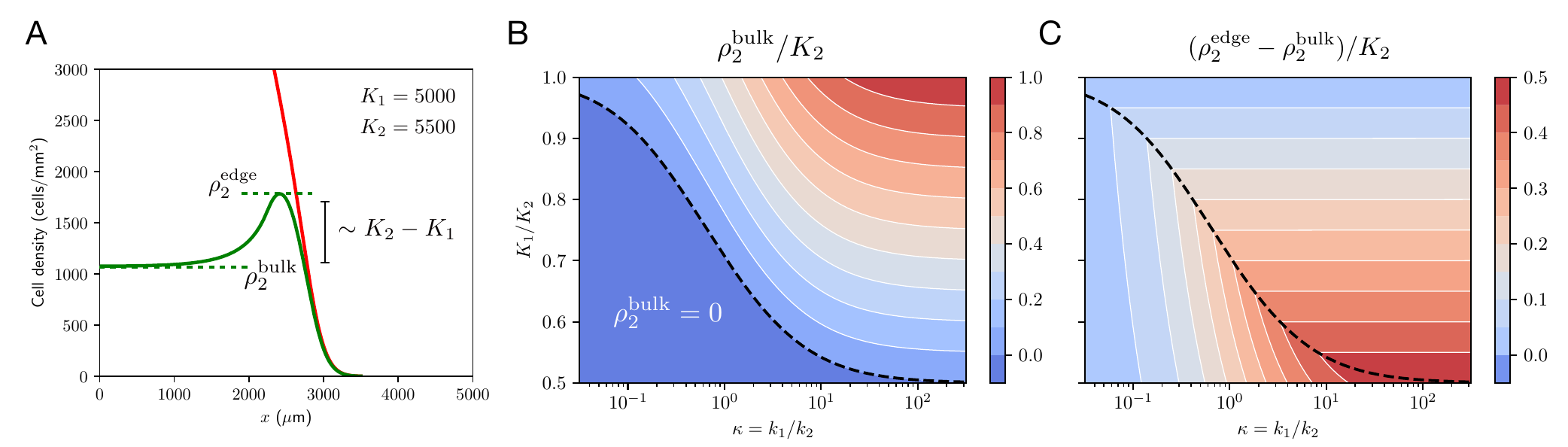}
    \caption{Cell cycle transition rates ($k_1,\,k_2$), and crowding constraints ($K_1,\, K_2$) determine cell proliferation patterns in growing tissues. (A) Schematic of travelling wave solutions near the tissue edge. (B)-(C) Approximated S/G2/M cell densities at the tissue edge and tissue bulk as a function of the ratios $\kappa = k_1/k_2$ and $K_1/K_2$. The black dashed line corresponds to the curve $K_1/K_2 = \alpha(\kappa) = 2/(\sqrt{\kappa^2+6\kappa+1}-\kappa-1)$.}
    \label{fig: figure 4}
\end{figure*}

The expression for the minimum travelling wave speed facilitates a comparison between the two-stage model proposed here (Eqs.~\eqref{eq: full model}), and conventional single-population models of cell migration of the form
\begin{equation*}
    \partial_t\rho= D\Delta \rho + r\rho F(\rho)\,,
\end{equation*}
where $F$ is a non-increasing function satisfying $F(0) = 1$.
The intrinsic growth rate of the population, $r$, is related to the intrinsic rates of cell cycle progression, $k_1$ and $k_2$, via $r^{-1} = k_1^{-1}+k_2^{-1}$. When $4r/(k_1+k_2)\ll 1$, Eq.~\eqref{eq: minimum travelling wave speed} can be approximated by 
\begin{equation*}
     c_{\text{min}}\sim 2\sqrt{Dr},\quad r = \frac{k_1k_2}{k_1+k_2}\,,
\end{equation*}
which agrees with the prediction of the well-known Fisher– Kolmogorov–Petrovsky–Piskunov (FKPP) equation ($F(\rho)=1-\rho/K$ for a maximum cellular density $K>0$) in one spatial dimension.  Using the estimated parameter values we obtain $4r/(k_1+k_2)\sim 0.98$, and in this case Eq.~\eqref{eq: minimum travelling wave speed} predicts a minimum travelling wave speed of $c_{\text{min}}\sim 33$ $\mu$m$/$h, while the FKPP approximation yields $c_{\text{min}}\sim 26$ $\mu$m$/$h; both of them within the measured values by Heinrich et al. \cite{heinrich2020size}.

A better comparison with the two population model can be obtained by setting $rF(\rho) = \lambda(\rho)$, where $\lambda(\rho)$ is the dominant eigenvalue of the growth matrix
\begin{align*}
\begin{pmatrix}
    -k_1 f(\rho) & 2k_2g(\rho)
    \\
    k_1 f(\rho) & -k_2g(\rho)
\end{pmatrix}\,,
\end{align*}
as given by Eqs.~\eqref{eq: full model}. In this case, 
\begin{align*}
    c_{\text{min}} = 2\sqrt{D\lambda(0)}\,,
\end{align*}
where $\lambda(0) = (-k_1-k_2+\sqrt{k_1^2+k_2^2+6k_1k_2})/2$, agreeing with the prediction from Eq.~\eqref{eq: minimum travelling wave speed}.

More generally, we noted that $c_{\text{min}}$ does not depend on the choice of \emph{crowding functions} $f$ and $g$; however, crowding constraints have an impact on the observed migration patterns (Fig.~\ref{fig: figure 3}B). To understand how growth and cell cycle regulation lead to the patterns observed experimentally, we investigate travelling wave solutions in a simplified version of our model, utilising the same parameters as in Eqs.~\eqref{eq: full model} and~\eqref{eq: crowding functions} (see Supplementary Information). In particular, we set $f(\rho) = H(K_1-\rho)$, and $g(\rho) = H(K_2-\rho)$, where $H(\cdot)$ denotes the Heaviside function. This reduced model does not accurately approximate the model presented in Eqs.~\eqref{eq: full model} and~\eqref{eq: crowding functions}, but nonetheless it captures the same qualitative behaviour, and hence we expect that the relevant phenomena show similar dependence with respect to model parameters (see Supplementary Fig. 4). The analysis of travelling wave solutions for this simpler model suggests that the density of S/G2/M cells in the tissue bulk, $\rho_2^{\text{bulk}}$, only depends on the ratio of cell cycle progression rates, $\kappa = k_1/k_2$, and on the ratio of densities associated with crowding constraints, $K_1/K_2$, (Fig.~\ref{fig: figure 4}). In particular, we obtain
\begin{equation*}
\frac{\rho_2^{\text{bulk}}}{K_2}\sim\!
\begin{dcases}
 \frac{K_1}{K_2}\frac{\sqrt{\kappa^2+6\kappa+1}-\kappa+1}{2}-1\,, & {K_1}/{K_2} > \alpha(\kappa) \,;
 \\
 0\,, & {K_1}/{K_2}\leq \alpha(\kappa)\,;
\end{dcases}
\end{equation*}
where $\alpha(\kappa) = 2/(\sqrt{\kappa^2+6\kappa+1}-\kappa-1)$. A similar dependence with respect to the model parameters is observed numerically for the model given by Eqs.~\eqref{eq: full model} and~\eqref{eq: crowding functions} (Supplementary Fig. 4). For our estimated parameters, the expression above predicts $\rho_2^{\text{bulk}}/K_2\sim 0.3$, which is consistent with experimental observations. We also highlight that, as long as $K_1<K_2$, and $k_1$ and $k_2$ are of a similar order of magnitude, this expression predicts that the number of cells in S/G2/M in the tissue bulk will be small in comparison to the number of cells in G1/post-M (Fig.~\ref{fig: figure 4}B). In particular, note that $\rho_2^{\text{bulk}}\rightarrow 0$ as $K_2\rightarrow \infty$.

Interestingly, the travelling wave analysis also reveals that, when $\rho_2^{\text{bulk}} > 0$, the difference in S/G2/M cell density between the tissue bulk, $\rho_2^{\text{bulk}}$,  and the tissue edge, $\rho_2^{\text{edge}}$, depends only on the difference of densities associated to crowding constraints at the G1-S and G2-M boundaries (Fig.~\ref{fig: figure 4}C),
 \begin{equation*}
\rho_2^{\text{edge}}-\rho_2^{\text{bulk}}\sim\!
\begin{dcases}
 K_2-K_1\,, & {K_1}/{K_2} > \alpha(\kappa) \,;
 \\
 {\rho_2^{\text{edge}}}\,, & {K_1}/{K_2}\leq \alpha(\kappa)\,;
\end{dcases}
\end{equation*}
where 
\begin{equation*}
\rho_2^{\text{edge}}\sim \frac{K_1}{K_2}\frac{\sqrt{\kappa^2+6\kappa+1}-\kappa-1}{2}\,.
\end{equation*}
For our estimated parameters, we obtain $\rho_2^{\text{edge}}-\rho_2^{\text{bulk}}\sim 500$ cells$/$mm$^2$, again consistent with the experimental observations. These analytical expressions confirm the impact of density-dependent effects on cell migration and suggest that differences in the regulation of cell cycle stages contribute to the emergence of cell proliferation patterns.

\subsection*{Density-dependent effects and experimental design}

\begin{SCfigure*}[.41]
\begin{wide}
    \includegraphics[width = .65\textwidth]{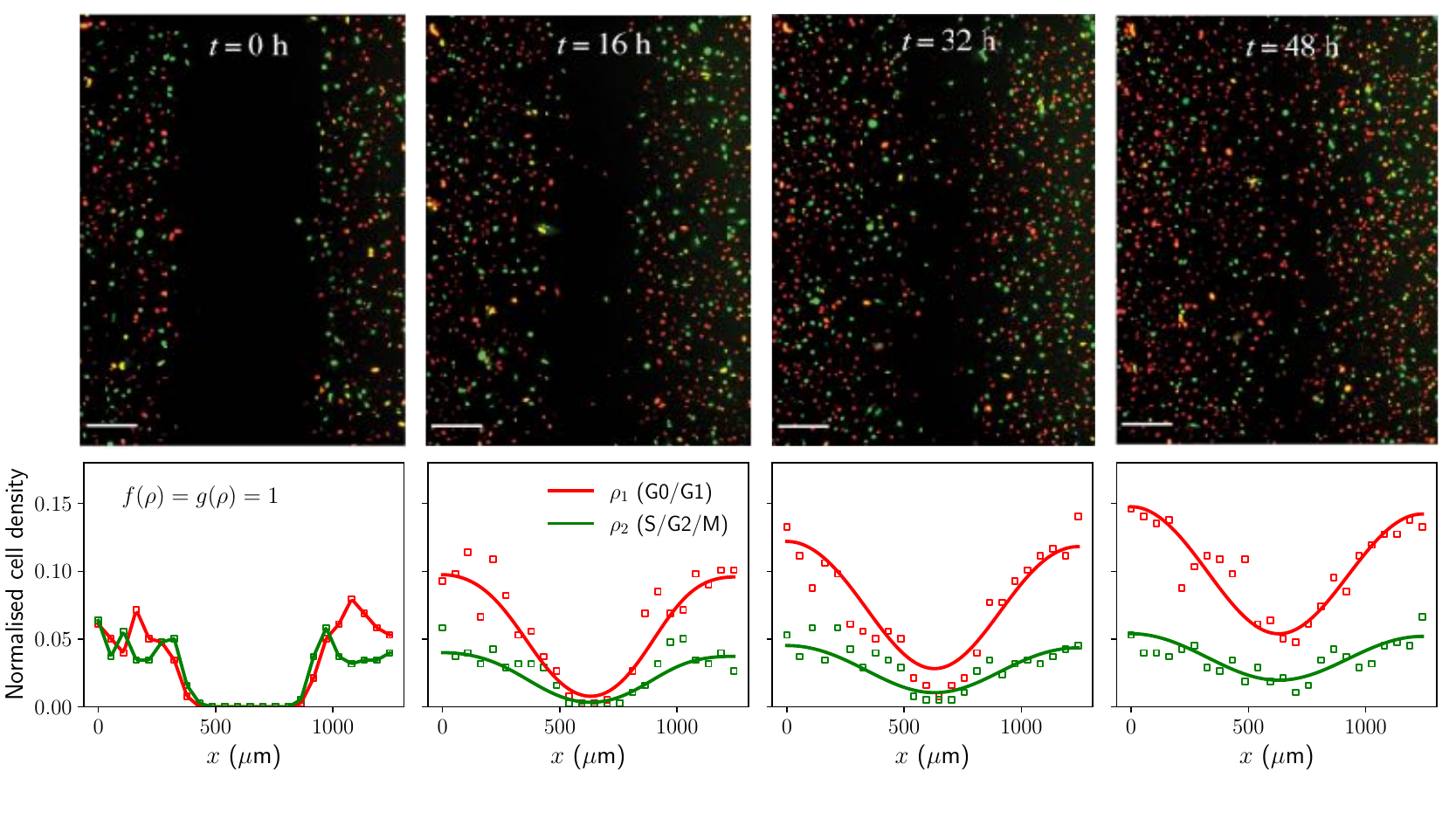}
    \caption{Absence of density-dependent effects in a low-density scratch-assay experiment (1205Lu melanoma cells). In this case, an exponential growth model can reproduce the experimental data. Density is normalised by using the theoretical maximum density corresponding to hexagonal close packing of cells \cite{vittadello2018mathematical}. Top row is adapted from \cite{simpson2020practical}, with scale bars corresponding to 200 $\mu$m. Numerical solutions of Eqs.~\eqref{eq: full model} with $f(\rho) = g(\rho)=1$ on a one-dimensional domain. Parameters are estimated using the experimental data from \cite{simpson2020practical} -- see Supplementary Fig. 6 for posterior distributions.}
    \label{fig: figure 5}
    \end{wide}
\end{SCfigure*}

We have demonstrated that our model can be calibrated to experimental data collected by Heinrich et al. \cite{heinrich2020size} to provide confident estimates of all parameters and, from there, used to extract and quantify crowding constraints regulating the cell cycle. An obvious question to ask is whether the model parameters could also be confidently estimated from other datasets, in particular where the cell density remains much lower and the impact of tissue crowding is reduced. To explore this question, we attempt to estimate the model parameters (including $K_1$ and $K_2$) using data from a low-density scratch assay with 1205Lu melanoma cells --- see Fig. \ref{fig: figure 5}. We highlight that melanoma cells are highly metastatic and often display uncontrolled and invasive migration, in contrast to the highly collective and regulated movement exhibited by epithelial cells. In this experiment, tissues are seeded at an initial density of $\sim 400$ cells$/$mm$^2$ ($5\%$ of the theoretical maximum packing density \cite{vittadello2018mathematical}), and data is collected every 16 hours, over two full days, allowing cells to potentially undergo 1-2 cell cycles. The posterior distributions obtained for the different model parameters reveal estimates for $D$, $k_1$ and $k_2$ that are consistent with previous studies \cite{simpson2020practical}. However, the low experimental densities do not allow for the quantification of density-dependent effects --- the parameters $K_1$ and $K_2$ cannot be estimated with any degree of confidence (see Supplementary Fig. 5). This non-identifiability of $K_1$ and $K_2$ suggests the use of a simpler model, which assumes that cell cycle progression is independent of density-dependent effects ($f(\rho) = g(\rho) = 1$) and hence is only valid in the low-density regime. Indeed, when calibrated to data from the low-density scratch assay it provides accurate parameters estimates (Supplementary Fig. 6), and an excellent agreement with the experimental data --- see Fig. \ref{fig: figure 5}. This result clearly illustrates both the key role that mathematical modelling can play in the experimental design process, and the importance of considering parameter identifiability in the process of model construction.

\section*{Discussion}

In this work, we have presented a new mathematical model of cell migration with cell cycle dynamics which captures and quantifies cell cycle regulation by sensing levels of tissue crowding. In line with previous experimental studies, by combining minimal modelling and Bayesian inference, we confirm that cell cycle progression is monitored via crowding constraints \cite{streichan2014spatial,donker2022mechanicalG2}, and present a systematic approach towards the quantification of interactions regulating cell proliferation. Our model is capable  of quantifying cell cycle data from experiments using the FUCCI system, and enables the extraction of mechanistic insights into how individual cells regulate proliferation based on population-level measures.

The model presented here offers several applications to further our understanding of cell-cell interactions in cell proliferation. In particular, our model presents a systematic way to quantify the impact of drugs and gene knockouts/knockdowns interfering with cell proliferation. By using parameter estimation techniques, applied to different experimental datasets, we can gain insights into the regulatory roles of specific genes in the cell cycle. Another possible application concerns the study of cell migration in biomaterials incorporating cadherin proteins, which have recently been shown to slow down cell cycle dynamics \cite{suh2023cadherin}. Furthermore, generalisations of the FUCCI system could allow for a finer representation of the different cell cycle stages -- for instance, FUCCI4 \cite{bajar2016fluorescent} allows for the simultaneous visualisation of the four stages of the cell cycle. In line with these methodologies, extensions of our model (Eqs.~\eqref{eq: full model}) to multi-stage cell populations are straightforward, and could enable a more exhaustive explanation of the role of spatial constraints across all four cell cycle stages \cite{donker2022mechanicalG2}.

In the case of Heinrich et al.'s experiments \cite{heinrich2020size}, the excellent imaging quality allowed us to perform an accurate quantification of the cellular density profiles. This, in turn, facilitated model development and the subsequent inference of model parameters from the data, with the estimated parameters showing a low uncertainty. While the parameter identifiability of such mathematical models can be evaluated a priori under the assumption of infinite ideal data \cite{renardy2022structural,browning2023structural}, biologically realistic datasets are finite, and often contain a significant level of noise, which can, in certain instances, constrain the ability to confidently estimate model parameters. More generally, and as we have illustrated, practical constraints in the experimental data often relate to the level of model complexity which can be inferred from experiments and the confidence in model parameter estimates.

Continuum models are a widely adopted approach for describing cell migration. However, these models come with limitations: they tend to neglect local structure, especially in situations involving multiple cell populations. Such local structure can be observed in Fig.~\ref{fig: figure 1}; (C) and (D) show some degree of local correlation in the cell phases, however this phenomenon is lost when averaging radially to obtain the density profiles in (E). Agent-based models \cite{klowss2022stochastic,hollring2023capturing, carpenter2024mechanical} can help mitigate some of these issues, by providing more understanding of the generation and maintenance of spatial structure, but at the cost of increased computational times for simulation and inference, additional model parameters, and limited analytical tractability. We emphasise, however, that cell cycle dynamics appear to be globally desynchronised, as observed in previous studies \cite{nowak2023impact}, and so our differential equation-based model remains appropriate for this study, where the data is generated by averaging over a number of experimental replicates. 

The model presented here is minimal in the sense that it assumes that cell movement is random, and it ignores basic cell-cell interactions which are typical of epithelial cell migration such as cell-cell adhesion. While local cell density is likely to have an impact on cell motility \cite{heinrich2020size}, previous work shows that for individual expanding epithelial tissues, the linear diffusion model provides a good approximation \cite{falco2023quantifying}. Note, however, that it is important to account for population pressure and its impact on cell movement when considering tissue-tissue interactions \cite{heinrich2021self}. Additional research is needed to determine whether more complicated models \cite{CarrilloMurakawaCellAdhesion,falco2022local}, incorporating cell-cell adhesion and other basic interactions offer deeper mechanistic understanding.
Moreover, the model given in Eqs.~\eqref{eq: full model} assumes that at low densities, the duration of each of the cell cycle stages follows an exponential distribution. While this assumption contradicts experimental observations \cite{smith1973cells,weber2014quantifying} and can be mitigated by representing the cell cycle as a multi-stage process \cite{yates2017multi, gavagnin2019invasion}, such models break the cell cycle into a very large number of stages, limiting the potential for calibration to experimental data. Additional investigation is required to understand the extent to which more complicated models can provide further insights into how cells coordinate proliferation and migration to give rise to complex collective behaviours. For example, in the context of the cell cycle, an option is to explicitly incorporate cell cycle stage via the use of an age-structured model \cite{kynaston2022equivalence} that includes density-dependent regulation. Our results indicate that adopting a quantitative approach \cite{liu2023parameter}, that carefully examines quantitative data through the lens of mathematical modelling and Bayesian inference, can help provide  answers to this question.

\section*{Authors' contributions}
CF and REB conceived the original
idea. CF created the code, carried out the analysis, and wrote the manuscript with input from REB. JAC helped supervise the project. DJC provided the experimental data and aided in interpreting the results. All authors gave final approval for publication.

\section*{Acknowledgments}
The authors would like to thank I. Breinyn for assistance with the experimental datasets. CF acknowledges support via a fellowship from "la Caixa" Foundation (ID 100010434) with code LCF/BQ/EU21/11890128. JAC was supported by the Advanced Grant Nonlocal-CPD (Nonlocal PDEs for Complex Particle Dynamics: Phase Transitions, Patterns and Synchronization) of the European Research Council Executive Agency (ERC) under the European Union’s Horizon 2020 research and innovation programme (grant agreement No. 883363).
JAC was also partially supported by EPSRC grants EP/T022132/1 and EP/V051121/1. REB and DJC would like to thank the Royal Society for an International Exchange Scheme grant. This work was also supported by a grant from the Simons Foundation (MP-SIP-00001828, REB).

\section*{Declaration of Interest}
The authors declare no competing interests.

\section*{Data and Code Availability}

Code to solve the model and to perform the parameter estimation is available on Github: \url{https://github.com/carlesfalco/InferenceCellCyclePDE}. Data used to calibrate the model can also be found on Github and in \cite{heinrich2020size}. Scratch assay data is taken from \cite{simpson2020practical}.

\clearpage



\section*{Supplementary material}

## S1   Bayesian parameter estimation

All experimental datasets [2, 8] consist of direct measurements of the density of cells in the G1/post-M, and S/G2/M phases of the cell cycle. We denote these measurements by $\{\rho_1^{\mathcal{D}}(\mathbf{x}_i, t_j), \rho_2^{\mathcal{D}}(\mathbf{x}_i, t_j)\}_{i,j}$. The next step in order to estimate the different model parameters is to assume a so-called error model, which relates experimental measurements with the model predictions given by the solutions of the model: $\rho_1(\mathbf{x}, t), \rho_2(\mathbf{x}, t)$. For simplicity, here we assume that the residuals are independent and normally distributed

$$\rho_1^{\mathcal{D}}(\mathbf{x}_i, t_j) - \rho_1(\mathbf{x}_i, t_j) \stackrel{\text{iid}}{\sim} N(0, \sigma_1^2),$$
$$\rho_2^{\mathcal{D}}(\mathbf{x}_i, t_j) - \rho_2(\mathbf{x}_i, t_j) \stackrel{\text{iid}}{\sim} N(0, \sigma_2^2),$$

where $\sigma_1$ and $\sigma_2$ are parameters to be estimated from the data.

The white noise assumption has the advantage that simple likelihood-based methods can be used for inference. In particular, the log-likelihood of observing the data, given specific model parameters $\theta$, can be written as

$$\ell_{\mathcal{D}}(\theta) = -\frac{1}{2} \sum_{k=1}^{2} \sum_{i,j} \left( \log \left(2\pi\sigma_k^2\right) + \left( \frac{\rho_k^{\mathcal{D}}(\mathbf{x}_i, t_j) - \rho_k(\mathbf{x}_i, t_j)}{\sigma_k} \right)^2 \right).$$

While we assume a simplistic noise model to perform the parameter inference, model misspecification and the temporal resolution of the measured data are likely to introduce correlations between residuals [3]. In particular, some degree of correlation might be expected given the model parameters are very well-determined with a relatively small variance – see Fig. 2 in the main text. More recently, a binomial measurement error model has been suggested in order to mitigate some of the inconsistencies of the white noise assumption [9]. Other commonly used error models assume different forms of multiplicative noise, which preserve the positivity of the data [5, 6]. A more comprehensive study of the error model is left as a subject for future investigation.

Maximising the log-likelihood function would give a set of parameters $\theta^*$ that we could use to generate further model predictions. However, this approach does not give any information on the associated uncertainty, which here is of particular interest given that the parameters are estimated from noisy experimental data. To explore parameter identifiability, we follow a Bayesian approach, in which uncertainty associated with model parameters ($\theta$) is quantified in a posterior distribution $P(\theta \,|\, \rho^{\mathcal{D}}) = P(\theta \,|\, \rho_1^{\mathcal{D}}, \rho_2^{\mathcal{D}})$. This posterior distribution can be

calculated from Bayes' theorem

$$P(\theta \,|\, \rho^{\mathcal{D}}) \propto P(\rho^{\mathcal{D}} \,|\, \theta) \pi(\theta),$$

where $P(\rho^{\mathcal{D}} \,|\, \theta) = \exp \ell_{\mathcal{D}}(\theta)$ is the likelihood of observing the measured data, and $\pi(\theta)$ is the prior distribution of the parameter vector $\theta$. For the tissue expansion experiments [2], we assume a log-uniform prior on $D, k_1, k_2$, with bounds: $10^1$ $\mu$m$^2$/h $< D <$ $10^4$ $\mu$m$^2$/h, $10^{-4}$ h$^{-1} < k_1, k_2 < 10^1$ h$^{-1}$. This assumption allows us to consider a broad range of orders of magnitude for these parameters, although simpler uniform priors could also be used. For the parameters $K_1, K_2, \sigma_1, \sigma_2$, uniform priors with the following bounds were used: $0 < K_1, K_2 < 20000$ cells/mm$^2$, $0 < \sigma_1, \sigma_2 < 2000$ cells/mm$^2$. For the scratch assay data, we follow [8] and assume $\sigma_1 = \sigma_2 = \sigma$, and a uniform prior in all model parameters with the following conservative bounds: $0 < D < 2000$ $\mu$m$^2$/h, $0 < k_1, k_2 < 0.2$ h$^{-1}$, $0 < K_1, K_2 < 30000$ cells/mm$^2$, $0 < \sigma < 4000$ cells/mm$^2$.

We use a Metropolis-Hastings MCMC (Markov chain Monte Carlo) sampler with adaptive proposal covariance to infer the posterior distributions. This is implemented in the parameter estimation toolbox pyPESTO [7]. In the MCMC algorithm, a Markov Chain starts at position $\theta$ and accepts a potential move to $\theta^*$ with probability $q = \min\{1, P(\theta \,|\, \rho^{\mathcal{D}})/P(\theta^* \,|\, \rho^{\mathcal{D}})\}$. In this way, the Markov chain tends to move towards high values of the posterior distribution, while still allowing for transitions to regions of lower probability in order to move away from local maxima. Figures S1 and S4 show typical MCMC iterations for both sets of experimental data and the corresponding data. We show the obtained stationary posterior distributions in the main text, and in Fig. S4.

## S2   Outline of the numerical scheme

We briefly explain the numerical scheme used to solve our model in polar coordinates. For the tissue expansion experiments we assume solutions with radial symmetry: $\rho_1(\mathbf{x}, t) = \rho_1(r, t)$, $\rho_2(\mathbf{x}, t) = \rho_2(r, t)$, where $r$ denotes the distance from the tissue centre. Hence, we can write

$$\Delta \rho_k = \partial_r^2 \rho_k + r^{-1} \partial_r \rho_k, \quad k = 1, 2\,.$$

We use a finite-volume scheme [2] and discretise the domain into a small circle $C_0$ of radius $r_{1/2} = \delta r/2$, and concentric annuli $C_i$ with inner radii $r_{i-1/2} = (i - 1/2)/\delta r$, for

$i = 1, 2, \ldots, N$. If $\mathbf{x} \in C_i$, we approximate

$$\rho_k(\mathbf{x}, t) \approx \rho_k^i(t) := \frac{1}{|C_i|} \int_{C_i} \rho_k, \quad i = 0, 1, \ldots, N;$$

where $|C_i|$ denotes the volume of $C_i$. In particular, by integrating the equation for $\rho_1$ over $C_0$ we obtain

$$\frac{\mathrm{d}\rho_1^0}{\mathrm{d}t} = \frac{2\pi D}{|C_0|} \int_0^{r_{1/2}} r \left( \partial_r^2 \rho_1 + r^{-1} \partial_r \rho_1 \right) \mathrm{d}r - \frac{k_1}{|C_0|} \int_{C_0} \rho_1 f(\rho) + \frac{2k_2}{|C_0|} \int_{C_0} \rho_2 g(\rho),$$

where $|C_0| = \pi r_{1/2}^2$ denotes the area of $C_0$. The first integral can be calculated exactly to obtain

$$\int_0^{r_{1/2}} r \left( \partial_r^2 \rho_1 + r^{-1} \partial_r \rho_1 \right) \mathrm{d}r = r (\partial_r \rho)\big|_{r=r_{1/2}}.$$

The last two integrals can be approximated to obtain

$$\frac{\mathrm{d}\rho_1^0}{\mathrm{d}t} = \frac{2\pi D}{|C_0|} r(\partial_r \rho_1)\big|_{r=r_{1/2}} - k_1 \rho_1^0 f(\rho^0) + 2k_2 \rho_2^0 g(\rho^0),$$

where $\rho^0 = \rho_1^0 + \rho_2^0$.

Similarly, we integrate the equation for $\rho_1$ over $C_i$, $i \geq 1$, to obtain

$$\frac{\mathrm{d}\rho_1^i}{\mathrm{d}t} = \frac{2\pi D}{|C_i|} \left( r(\partial_r \rho_1)\big|_{r=r_{i+1/2}} - r(\partial_r \rho_1)\big|_{r=r_{i-1/2}} \right) - k_1 \rho_1^i f(\rho^i) + 2k_2 \rho_2^i g(\rho^i),$$

where $|C_i| = \pi(r_{i+1/2}^2 - r_{i-1/2}^2)$ and $\rho^i = \rho_1^i + \rho_2^i$. An analogous set of equations can be obtained for $\rho_2$ following the same arguments. Finally, we approximate the derivatives $\partial_r \rho$ as

$$(\partial_r \rho_k)_{r=r_{i+1/2}} \approx \frac{\rho_k^{i+1} - \rho_k^i}{\delta r}, \quad k = 1, 2; \; i = 0, 1, \ldots, N.$$

We solve the resulting differential equations using a fourth-order Runge-Kutta method implemented in the `scipy.integrate.ode` class in Python.

## S3 Minimum travelling wave speed

We look for travelling solutions in the model given by Eqs. (2) in the main text, in one spatial dimension. We assume that the *crowding functions* $f(\rho)$ and $g(\rho)$ are non-increasing with $\rho$, and non-negative. In the comoving reference frame, we can write: $\rho_1(x,t) = U_1(z)$, $\rho_2(x,t) = U_2(z)$, where $z = x - ct$, and $c \geq 0$ denotes the wave speed. By denoting $U = U_1 + U_2$, $V_1 = U_1'$, $V_2 = U_2'$, the model reduces to

$$\begin{cases} U_1' &= V_1\,, \\ DV_1' &= -cV_1 + k_1 U_1 f(U) - 2k_2 U_2 g(U)\,, \\ U_2' &= V_2\,, \\ DV_2' &= -cV_2 - k_1 U_1 f(U) + k_2 U_2 g(U)\,, \end{cases} \quad (1)$$

where the primes indicate differentiation with respect to $z$.

The set of steady states of system (1) consists of the origin $(U_1, V_1, U_2, V_2) = (0,0,0,0)$ and any state of the form $(\alpha, 0, U^* - \alpha, 0)$, with $f(U^*) = g(U^*) = 0$ and $0 \leq \alpha \leq U^*$. Note that whenever $f(\rho), g(\rho) > 0$ for all $\rho \geq 0$, the latter does not exist. As usual with linear diffusion models, the stability of the origin gives a lower bound on the wave speed $c$. In particular the Jacobian of system (1) at the origin reads

$$\begin{pmatrix} 0 & 1 & 0 & 0 \\ k_1/D & -c/D & -2k_2/D & 0 \\ 0 & 0 & 0 & 1 \\ -k_1/D & 0 & k_2/D & -c/D \end{pmatrix}.$$

The eigenvalues $\lambda_i$ of the linearized system about this point satisfy the polynomial equation

$$\lambda^4 + \frac{2c}{D}\lambda^3 + \left(\left(\frac{c}{D}\right)^2 - \frac{k_1+k_2}{D}\right)\lambda^2 - c\frac{k_1+k_2}{D^2}\lambda - k_1 k_2 = 0\,.$$

By defining

$$\gamma^{\pm} = \left(\frac{c}{D}\right)^2 + \frac{2}{D}\left[k_1 + k_2 \pm \sqrt{k_1^2 + k_2^2 + 6k_1 k_2}\right],$$

the roots of this quartic polynomial can be expressed as

$$\lambda_1^{\pm} = \frac{1}{2}\left(-\frac{c}{D} \pm \sqrt{\gamma^+}\right), \quad \lambda_2^{\pm} = \frac{1}{2}\left(-\frac{c}{D} \pm \sqrt{\gamma^-}\right).$$

5

We seek biologically realistic solutions with $U_1, U_2 \geq 0$, and hence the eigenvalues must be real. In particular, this demands $\gamma^{\pm} \geq 0$, which establishes the minimum travelling wave speed found in [10]

$$c_{\min} = \sqrt{2D\left(-k_1 - k_2 + \sqrt{k_1^2 + k_2^2 + 6k_1 k_2}\right)}. \qquad (2)$$

By writing

$$k_1^2 + k_2^2 + 6k_1 k_2 = (k_1 + k_2)^2 + 4k_1 k_2 = (k_1 + k_2)^2 \left[1 + \frac{4k_1 k_2}{(k_1 + k_2)^2}\right],$$

we observe that when $4k_1 k_2/(k_1 + k_2)^2 \ll 1$, the minimum travelling wave speed can be approximated by

$$c_{\min} \approx 2\sqrt{D\frac{k_1 k_2}{k_1 + k_2}},$$

which agrees with the minimum speed predicted by the Fisher–Kolmogorov–Petrovsky–Piskunov (FKPP) equation [4].

## S4  Study of travelling wave solutions

A commonly used approach to obtain approximate solutions for travelling waves is the so-called Canosa's method [1]. This procedure is a standard singular perturbation technique, and consists of a transformation $y = -z/c$, where $D/c^2 := \varepsilon$ is treated as a small parameter. The first-order perturbation in $\varepsilon$ approximates, within a small error, travelling solutions of the well-known FKPP equation, even though in this case $\varepsilon$ is not necessarily small [4]. In our case, by using the esimated parameters and a wave speed of 30 $\mu$m/h, we obtain $\varepsilon \sim O(1)$. We highlight, however, that the lowest order approximation in $\varepsilon$ provides an excellent approximation of the travelling wave – see Fig. 1.

By using the transformation $y = -z/c$, we can write system (1) as

$$\frac{dU_1}{dy} - \varepsilon \frac{d^2 U_1}{dy^2} + k_1 U_1 f(U) - 2k_2 U_2 g(U) = 0, \qquad (3)$$

$$\frac{dU_2}{dy} - \varepsilon \frac{d^2 U_2}{dy^2} - k_1 U_1 f(U) + k_2 U_2 g(U) = 0. \qquad (4)$$

Observe that, given the sign of the transformation $y = -z/c$, we need to impose the following

6

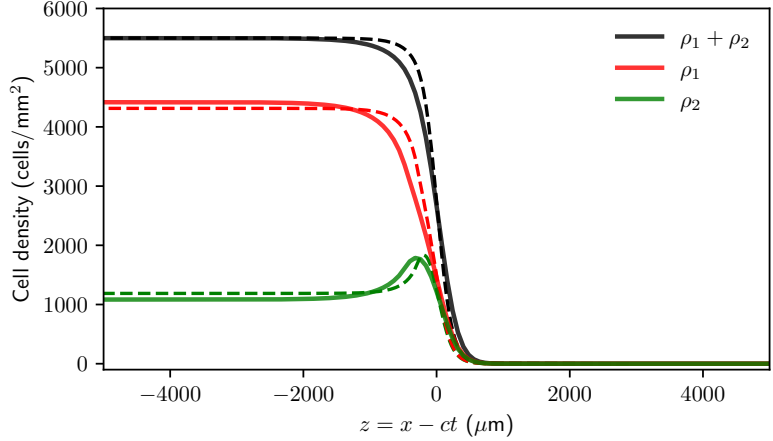

**Figure 1:** Comparison of travelling wave solutions obtained from the partial differential model (solid lines), given by Eqs. (1) in the main text, and the order $O(1)$ approximation (dashed lines), obtained from solving the ordinary differential equations (5) and (6). Model parameters are taken from posterior distribution modes.

boundary conditions

$$U_1(-\infty) = U_2(-\infty) = 0, \quad U_1(+\infty) = \alpha, \quad U_2(+\infty) = U^* - \alpha,$$

with $f(U^*) = g(U^*) = 0$ and $0 \leq \alpha \leq U^*$. For the choice of $f$ and $g$ in the main text ($f(U) = (1 - U/K_1)_+$ and $g(U) = (1 - U/K_2)_+$ with $K_1 < K_2$), we expect $U^* = K_2$ and $\alpha \in [0, K_2]$.

Although the analysis as $\varepsilon \to 0$ looks like a singular perturbation problem, setting $\varepsilon = 0$ gives a valid first-order approximation. This is due to the fact that the nonlinear terms in Eqs. (3) and (4) vanish at both boundaries [4]. Hence, we can look for a regular perturbation expansion in both $U_1$ and $U_2$. By denoting the order $O(1)$ solutions as $u_1$ and $u_2$ we obtain

$$\frac{\mathrm{d}u_1}{\mathrm{d}y} = -k_1 u_1 f(u) + 2k_2 u_2 g(u), \tag{5}$$

$$\frac{\mathrm{d}u_2}{\mathrm{d}y} = k_1 u_1 f(u) - k_2 u_2 g(u), \tag{6}$$

where $u = u_1 + u_2$. In the figure above (Fig. 1), we compare the approximate solutions obtained by solving this system with the full travelling wave solutions. We highlight that the lowest order approximation provides an excellent approximation of the travelling wave shape.

## S4.1 A simplified model

In order to make analytical progress we set $f$ and $g$ to be Heaviside functions: $f(u) = H(K_1-u)$ and $g(u) = H(K_2-u)$. This model is not an approximation of the model presented in the main text, but a simplification which preserves the same qualitative behaviour. Hence, we expect that the observed phenomena show similar dependence on the model parameters; this will be numerically confirmed later. In particular, note that this simplified model also describes two density checkpoints, at the G1-S boundary, and during the G2/M phases. The parameters $K_1$ and $K_2$, respectively, quantify the cell density associated with these checkpoints.

We rewrite Eqs. (5) and. (6) in terms of the variables $(u, u_2)$,

$$\frac{\mathrm{d}u}{\mathrm{d}y} = k_2 u_2 g(u),$$
$$\frac{\mathrm{d}u_2}{\mathrm{d}y} = k_1(u - u_2)f(u) - k_2 u_2 g(u).$$

Depending on the relative values of the total cell density, $u$, and the density checkpoints parameters $K_1$, $K_2$, we distinguish three possible cases. As inferred from the experimental data, we assume $K_1 < K_2$.

<u>Tissue edge ($u < K_1 < K_2$)</u>. In this region $f(u) = g(u) = 1$ and we can write

$$\frac{\mathrm{d}u}{\mathrm{d}y} = k_2 u_2,$$
$$\frac{\mathrm{d}u_2}{\mathrm{d}y} = k_1 u - (k_1 + k_2)u_2.$$

The solution at the tissue edge reads

$$u(y) = e^{-(k_1+k_2)y/2}\left[Ae^{\alpha y/2} + Be^{-\alpha y/2}\right],$$
$$u_2(y) = e^{-(k_1+k_2)y/2}\left[\frac{A(\alpha - (k_1+k_2))}{2k_2}e^{\alpha y/2} - \frac{B(\alpha + (k_1+k_2))}{2k_2}e^{-\alpha y/2}\right],$$

where $A, B$ are constants to be determined, and $\alpha = \sqrt{(k_1+k_2)^2 + 4k_1 k_2}$. Imposing boundary conditions at $y \longrightarrow -\infty$, and noting that $\alpha - (k_1+k_2) > 0$, we obtain $B = 0$. Hence, for $u < K_1$ both solutions are increasing exponentials. Without loss of generality we set

$U(0) = K_1$, giving $A = K_1$, and thus

$$u(y) = K_1 e^{(\alpha - k_1 - k_2)y/2},$$
$$u_2(y) = \frac{K_1 \left(\alpha - (k_1 + k_2)\right)}{2k_2} e^{(\alpha - k_1 - k_2)y/2}. \tag{7}$$

<u>Intermediate region $(K_1 < u < K_2)$.</u> In this region $f(u) = 0$ and $g(u) = 1$, leading to

$$\frac{\mathrm{d}u}{\mathrm{d}y} = k_2 u_2,$$
$$\frac{\mathrm{d}u_2}{\mathrm{d}y} = -k_2 u_2.$$

Hence, $u_2(y) = C e^{-k_2 y}$ for a constant $C$, which can be found by continuity at $z = 0$. We obtain

$$u_2(y) = \frac{K_1 \left(\alpha - (k_1 + k_2)\right)}{2k_2} e^{-k_2 y}. \tag{8}$$

Given that this is a decreasing exponential, we have found that the peak in S/G2/M cell density, $\rho_2^{\text{edge}}$, occurs at $y = 0$. This is, $\rho_2^{\text{edge}} = u_2(0)$.

Since in this region $u + u_2$ is a constant, which can be found by continuity at $y = 0$, we also obtain

$$u(y) = K_1 + u_2(0) - u_2(y).$$

<u>Tissue bulk $(K_1 < K_2 < u)$.</u> Now we have $f(u) = g(u) = 0$ and hence we can write

$$\frac{\mathrm{d}u}{\mathrm{d}y} = \frac{\mathrm{d}u_2}{\mathrm{d}y} = 0.$$

In the tissue bulk, both densities are constant and $u = K_2$. By using continuity, and the solutions from the intermediate region, we find

$$\rho_2^{\text{bulk}} := u_2(z) = (K_1 - K_2 + u_2(0))_+, \tag{9}$$

where we impose positivity of $u_2(z)$.

In particular, we find that, whenever $\rho_2^{\text{bulk}} > 0$, the S/G2/M cell density difference

9

between the tissue edge and the bulk satisfies

$$\rho_2^{\text{edge}} - \rho_2^{\text{bulk}} = K_2 - K_1 \,.$$

In this case, and by combining Eqs. (7), (8), and (9), we obtain the full solution

$$u_2(y) = \begin{cases} \frac{K_1(\alpha - (k_1 + k_2))}{2k_2} e^{(\alpha - k_1 - k_2)y/2}, & y \leq 0 \,; \\ \frac{K_1(\alpha - (k_1 + k_2))}{2k_2} e^{-k_2 y}, & 0 < y \leq y^* \,; \\ \frac{K_1(\alpha - (k_1 + k_2))}{2k_2} - (K_2 - K_1), & y > y^* \,, \end{cases} \quad (10)$$

for $y = -(x - ct)/c$, and $y^*$ defined from Eq. (8): $k_2 y^* = \log(\rho_2^{\text{edge}}/\rho_2^{\text{bulk}})$.

# S5 Supplementary Figures

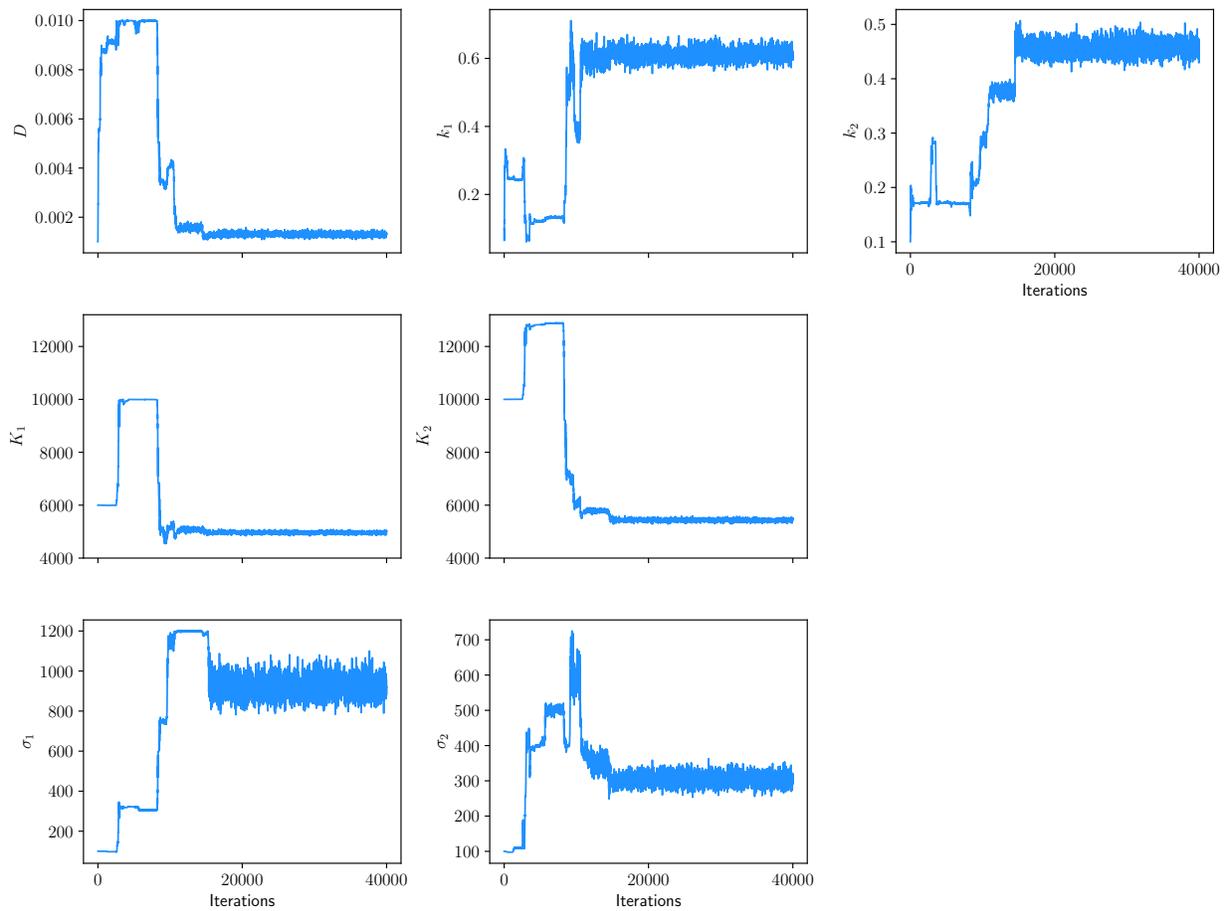

**Figure S1:** MCMC iterations for the large tissue expansions experimental data. Parameters $D, k_1, k_2, K_1, K_2$ correspond to the model presented in the main text, and $\sigma_1, \sigma_2$ are error model parameters.

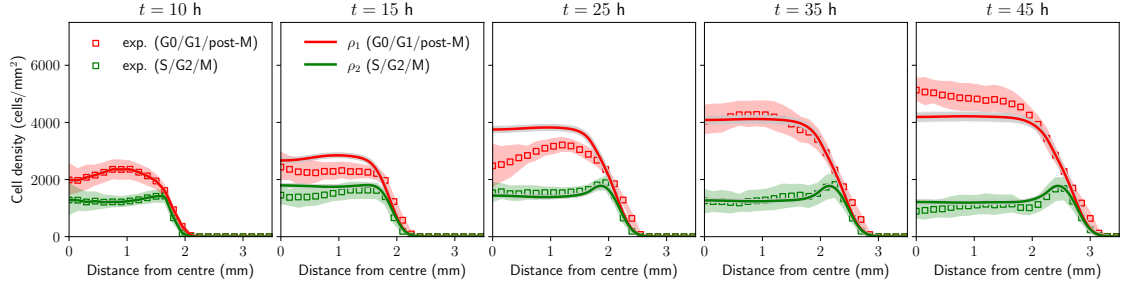

**Figure S2:** Comparing data (squares) and model predictions (solid lines) for large tissue expansions. Coloured shaded regions denote one experimental standard deviation with respect to the mean, obtained by averaging eleven experimental realisations. Gray shaded regions represent 95% confidence intervals obtained from the posterior distributions.

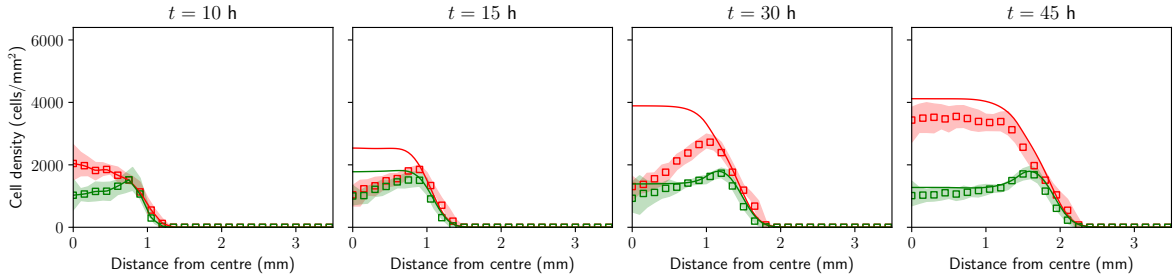

**Figure S3:** Comparing data (squares) and model predictions (solid lines) for small tissue expansions. Shaded regions denote one standard deviation with respect to the mean, obtained by averaging five experimental realisations. Numerical simulations in polar coordinates were obtained by using the parameter values obtained from the large tissue expansions, and no-flux boundary conditions. In order to minimise the effects of the stencil removal on cell behaviour, the initial condition corresponds to the experimental density profile ten hours after stencil removal.

12

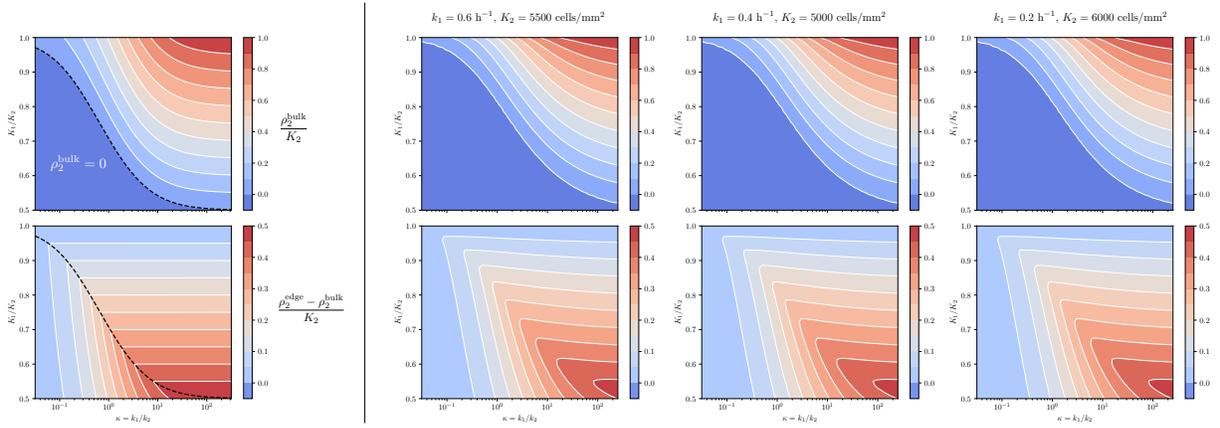

**Figure S4:** S/G2/M cell densities at the tissue edge and tissue bulk as a function of the ratios $\kappa = k_1/k_2$ and $K_1/K_2$: comparison between the simplified model (left) and the model presented in the main text (right). For the model with $f(\rho) = (1 - \rho/K_1)_+$, $g(\rho) = (1 - \rho/K_2)_+$, plotted values are obtained by solving numerically Eqs. (5) and (6) with different parameter values, as indicated in the figure. These results confirm that $\rho_2^{\text{edge}}$ and $\rho_2^{\text{bulk}}$ are determined by the two ratios of parameters: $k_1/k_2$ and $K_1/K_2$.

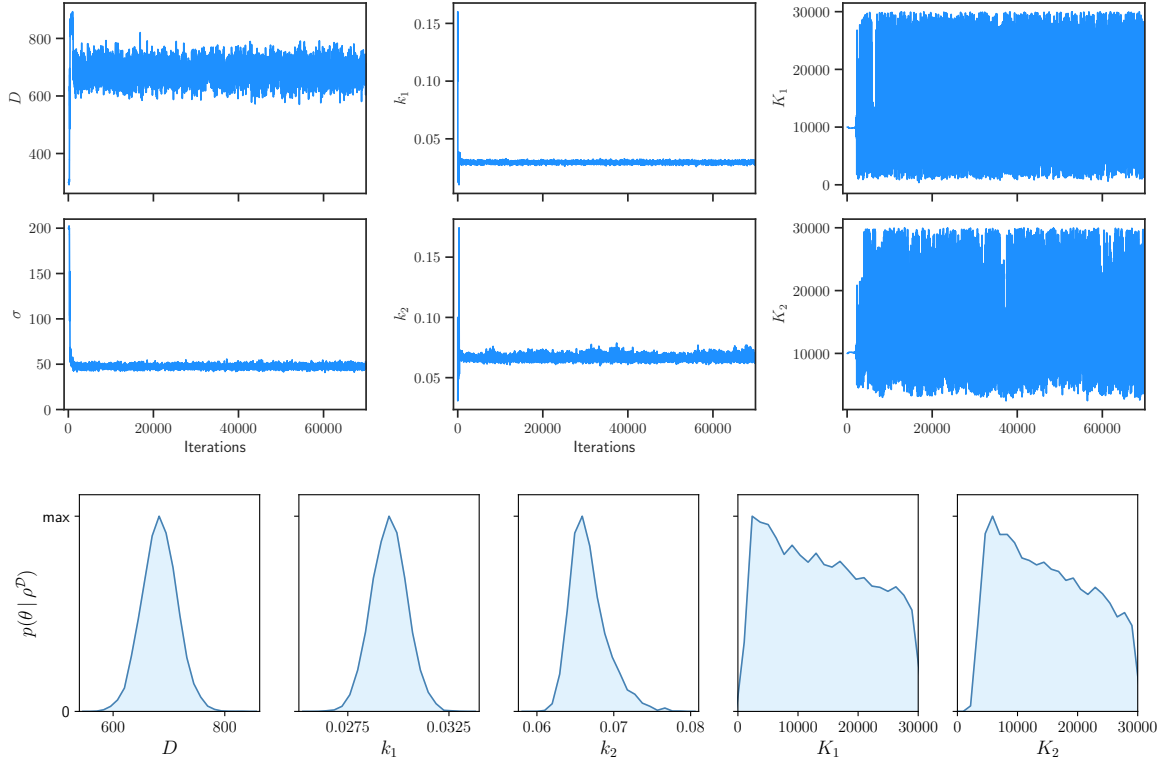

**Figure S5:** MCMC iterations and univariate marginal posterior distributions obtained by using low-density scratch assay data [8]. With this dataset, $K_1$ and $K_2$ are practically non-identifiable.

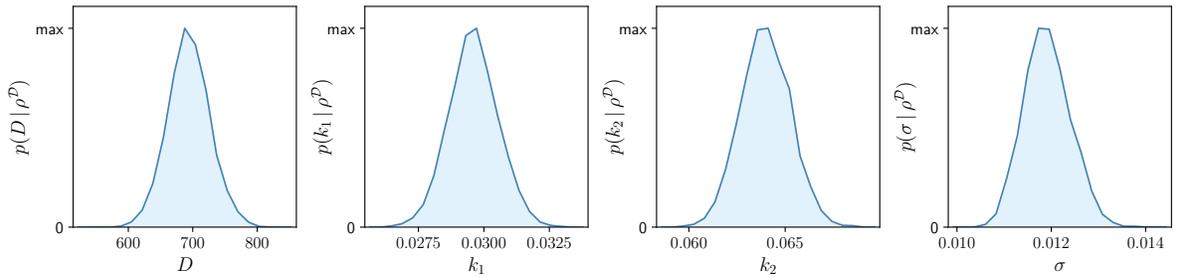

**Figure S6:** Univariate marginal posterior distributions for the exponential growth model ($f(\rho) = g(\rho) = 1$) obtained by using low-density scratch assay data [8]. In this case, the model has three parameters ($D, k_1, k_2$) and one error model parameter ($\sigma$, normalised by the theoretical maximum density assuming hexagonal packing of cells [10]).

14




\begin{thebibliography}{10}

\bibitem{JORGENSEN2004R1014}
P.~Jorgensen and M.~Tyers.
\newblock How cells coordinate growth and division.
\newblock {\em Current Biology}, 14(23):R1014--R1027, 2004.

\bibitem{streichan2014spatial}
S.~J. Streichan, C.~R. Hoerner, T.~Schneidt, D.~Holzer, and L.~Hufnagel.
\newblock Spatial constraints control cell proliferation in tissues.
\newblock {\em Proceedings of the National Academy of Sciences},
  111(15):5586--5591, 2014.

\bibitem{massague2004g1}
J.~Massagu{\'e}.
\newblock G1 cell-cycle control and cancer.
\newblock {\em Nature}, 432(7015):298--306, 2004.

\bibitem{MCCLATCHEY2012685}
A.~I. McClatchey and A.~S. Yap.
\newblock Contact inhibition (of proliferation) redux.
\newblock {\em Current Opinion in Cell Biology}, 24(5):685--694, 2012.

\bibitem{otto2017cell}
T.~Otto and P.~Sicinski.
\newblock Cell cycle proteins as promising targets in cancer therapy.
\newblock {\em Nature Reviews Cancer}, 17(2):93--115, 2017.

\bibitem{gupta2022mechanical}
V.~K. Gupta and O.~Chaudhuri.
\newblock Mechanical regulation of cell-cycle progression and division.
\newblock {\em Trends in Cell Biology}, 32(9):773--785, 2022.

\bibitem{vittadello2018mathematical}
S.~T. Vittadello, S.~W. McCue, G.~Gunasingh, N.~K. Haass, and M.~J. Simpson.
\newblock Mathematical models for cell migration with real-time cell cycle
  dynamics.
\newblock {\em Biophysical Journal}, 114(5):1241--1253, 2018.

\bibitem{simpson2020practical}
M.~J. Simpson, R.~E. Baker, S.~T. Vittadello, and O.~J. Maclaren.
\newblock Practical parameter identifiability for spatio-temporal models of
  cell invasion.
\newblock {\em Journal of the Royal Society Interface}, 17(164):20200055, 2020.

\bibitem{gavagnin2019invasion}
E.~Gavagnin, M.~J. Ford, R.~L. Mort, T.~Rogers, and C.~A. Yates.
\newblock The invasion speed of cell migration models with realistic cell cycle
  time distributions.
\newblock {\em Journal of Theoretical Biology}, 481:91--99, 2019.

\bibitem{sakaue2008visualizing}
A.~Sakaue-Sawano, H.~Kurokawa, T.~Morimura, A.~Hanyu, H.~Hama, H.~Osawa,
  S.~Kashiwagi, K.~Fukami, T.~Miyata, H.~Miyoshi, et~al.
\newblock Visualizing spatiotemporal dynamics of multicellular cell-cycle
  progression.
\newblock {\em Cell}, 132(3):487--498, 2008.

\bibitem{ridenour2012cycletrak}
D.~A. Ridenour, M.~C. McKinney, C.~M. Bailey, and P.~M. Kulesa.
\newblock Cycletrak: a novel system for the semi-automated analysis of cell
  cycle dynamics.
\newblock {\em Developmental Biology}, 365(1):189--195, 2012.

\bibitem{bajar2016fluorescent}
B.~T. Bajar, A.~J. Lam, R.~K. Badiee, Y.-H. Oh, J.~Chu, X.~X. Zhou, N.~Kim,
  B.~B. Kim, M.~Chung, A.~L. Yablonovitch, et~al.
\newblock Fluorescent indicators for simultaneous reporting of all four cell
  cycle phases.
\newblock {\em Nature Methods}, 13(12):993--996, 2016.

\bibitem{uroz2018regulation}
M.~Uroz, S.~Wistorf, X.~Serra-Picamal, V.~Conte, M.~Sales-Pardo,
  P.~Roca-Cusachs, R.~Guimer{\`a}, and X.~Trepat.
\newblock Regulation of cell cycle progression by cell-cell and cell-matrix
  forces.
\newblock {\em Nature Cell Biology}, 20(6):646--654, 2018.

\bibitem{pardee1989g1}
A.~B. Pardee.
\newblock {G1 events and regulation of cell proliferation}.
\newblock {\em Science}, 246(4930):603--608, 1989.

\bibitem{godard2019cell}
B.~G. Godard and C.-P. Heisenberg.
\newblock Cell division and tissue mechanics.
\newblock {\em Current Opinion in Cell Biology}, 60:114--120, 2019.

\bibitem{mckeown2019nutrient}
C.~R. McKeown and H.~T. Cline.
\newblock {Nutrient restriction causes reversible G2 arrest in Xenopus neural
  progenitors}.
\newblock {\em Development}, 146(20):dev178871, 2019.

\bibitem{CELORA2022111104}
G.~L. Celora, S.~B. Bader, E.~M. Hammond, P.~K. Maini, J.~M. Pitt-Francis, and
  H.~M. Byrne.
\newblock {A DNA-structured mathematical model of cell-cycle progression in
  cyclic hypoxia}.
\newblock {\em Journal of Theoretical Biology}, 545:111104, 2022.

\bibitem{donker2022mechanicalG2}
L.~Donker, R.~Houtekamer, M.~Vliem, F.~Sipieter, H.~Canever,
  M.~Gómez-González, M.~Bosch-Padrós, W.-J. Pannekoek, X.~Trepat, N.~Borghi,
  and M.~Gloerich.
\newblock {A mechanical G2 checkpoint controls epithelial cell division through
  E-cadherin-mediated regulation of Wee1-Cdk1}.
\newblock {\em Cell Reports}, 41(2):111475, 2022.

\bibitem{heinrich2020size}
M.~A. Heinrich, R.~Alert, J.~M. LaChance, T.~J. Zajdel, A.~Ko{\v{s}}mrlj, and
  D.~J. Cohen.
\newblock Size-dependent patterns of cell proliferation and migration in
  freely-expanding epithelia.
\newblock {\em eLife}, 9:e58945, 2020.

\bibitem{suh2023cadherin}
K.~Suh, Y.~K. Cho, I.~B. Breinyn, and D.~J. Cohen.
\newblock {E-cadherin biointerfaces reprogram collective cell migration and
  cell cycling by forcing homeostatic conditions}.
\newblock {\em bioRxiv 2023.07.25.550505}, 2023.

\bibitem{hollring2023capturing}
K.~H{\"o}llring, L.~Nui{\'c}, L.~Rogi{\'c}, S.~Kaliman, S.~Gehrer, C.~Wollnik,
  F.~Rehfeldt, M.~Hubert, and A.-S. Smith.
\newblock Capturing the mechanosensitivity of cell proliferation in models of
  epithelium.
\newblock {\em bioRxiv 2023.01.31.526438}, 2023.

\bibitem{vittadello2020examining}
S.~T. Vittadello, S.~W. McCue, G.~Gunasingh, N.~K. Haass, and M.~J. Simpson.
\newblock Examining go-or-grow using fluorescent cell-cycle indicators and
  cell-cycle-inhibiting drugs.
\newblock {\em Biophysical Journal}, 118(6):1243--1247, 2020.

\bibitem{Hines2014DeterminationOP}
K.~E. Hines, T.~R. Middendorf, and R.~W. Aldrich.
\newblock {Determination of parameter identifiability in nonlinear biophysical
  models: A Bayesian approach}.
\newblock {\em The Journal of General Physiology}, 143(3):401--416, 2014.

\bibitem{falco2023quantifying}
C.~Falc{\'o}, D.~J. Cohen, J.~A. Carrillo, and R.~E. Baker.
\newblock {Quantifying tissue growth, shape and collision via continuum models
  and Bayesian inference}.
\newblock {\em Journal of the Royal Society Interface}, 20(204):20230184, 2023.

\bibitem{pypesto}
Y.~Schälte, F.~Fröhlich, P.~J. Jost, J.~Vanhoefer, D.~Pathirana, P.~Stapor,
  P.~Lakrisenko, D.~Wang, E.~Raimúndez, S.~Merkt, L.~Schmiester, P.~Städter,
  S.~Grein, E.~Dudkin, D.~Doresic, D.~Weindl, and J.~Hasenauer.
\newblock {pyPESTO: a modular and scalable tool for parameter estimation for
  dynamic models}.
\newblock {\em Bioinformatics}, 39(11):btad711, 2023.

\bibitem{lachance2020practical}
J.~LaChance and D.~J. Cohen.
\newblock Practical fluorescence reconstruction microscopy for large samples
  and low-magnification imaging.
\newblock {\em PLoS Computational Biology}, 16(12):e1008443, 2020.

\bibitem{puliafito2012collective}
A.~Puliafito, L.~Hufnagel, P.~Neveu, S.~Streichan, A.~Sigal, D.~K. Fygenson,
  and B.~I. Shraiman.
\newblock Collective and single cell behavior in epithelial contact inhibition.
\newblock {\em Proceedings of the National Academy of Sciences},
  109(3):739--744, 2012.

\bibitem{warne2017optimal}
D.~J. Warne, R.~E. Baker, and M.~J. Simpson.
\newblock Optimal quantification of contact inhibition in cell populations.
\newblock {\em Biophysical Journal}, 113(9):1920--1924, 2017.

\bibitem{JIN2016136}
W.~Jin, E.~T. Shah, C.~J. Penington, S.~W. McCue, L.~K. Chopin, and M.~J.
  Simpson.
\newblock Reproducibility of scratch assays is affected by the initial degree
  of confluence: Experiments, modelling and model selection.
\newblock {\em Journal of Theoretical Biology}, 390:136--145, 2016.

\bibitem{jin2017logistic}
W.~Jin, E.~T. Shah, C.~J. Penington, S.~W. McCue, P.~K. Maini, and M.~J.
  Simpson.
\newblock Logistic proliferation of cells in scratch assays is delayed.
\newblock {\em Bulletin of Mathematical Biology}, 79(5):1028--1050, 2017.

\bibitem{renardy2022structural}
M.~Renardy, D.~Kirschner, and M.~Eisenberg.
\newblock {Structural identifiability analysis of age-structured PDE epidemic
  models}.
\newblock {\em Journal of Mathematical Biology}, 84(1):1--30, 2022.

\bibitem{browning2023structural}
A.~P. Browning, M.~Ta{\c{s}}c{\u{a}}, C.~Falc{\'o}, and R.~E. Baker.
\newblock Structural identifiability analysis of linear
  reaction--advection--diffusion processes in mathematical biology.
\newblock {\em Proceedings of the Royal Society A}, 480(2286):20230911, 2024.

\bibitem{klowss2022stochastic}
J.~J. Klowss, A.~P. Browning, R.~J. Murphy, E.~J. Carr, M.~J. Plank,
  G.~Gunasingh, N.~K. Haass, and M.~J. Simpson.
\newblock A stochastic mathematical model of 4d tumour spheroids with real-time
  fluorescent cell cycle labelling.
\newblock {\em Journal of the Royal Society Interface}, 19(189):20210903, 2022.

\bibitem{carpenter2024mechanical}
L.~C. Carpenter, F.~P{\'e}rez-Verdugo, and S.~Banerjee.
\newblock Mechanical control of cell proliferation patterns in growing
  epithelial monolayers.
\newblock {\em Biophysical Journal}, 123(7):909--919, 2024.

\bibitem{nowak2023impact}
C.~M. Nowak, T.~Quarton, and L.~Bleris.
\newblock Impact of variability in cell cycle periodicity on cell population
  dynamics.
\newblock {\em PLOS Computational Biology}, 19(6):e1011080, 2023.

\bibitem{heinrich2021self}
M.~A. Heinrich, R.~Alert, A.~E. Wolf, A.~Ko{\v{s}}mrlj, and D.~J. Cohen.
\newblock Self-assembly of tessellated tissue sheets by expansion and
  collision.
\newblock {\em Nature Communications}, 13(1):1--10, 2022.

\bibitem{CarrilloMurakawaCellAdhesion}
J.~A. Carrillo, H.~Murakawa, M.~Sato, H.~Togashi, and O.~Trush.
\newblock A population dynamics model of cell-cell adhesion incorporating
  population pressure and density saturation.
\newblock {\em Journal of Theoretical Biology}, 474:14--24, 2019.

\bibitem{falco2022local}
C.~Falc{\'o}, R.~E. Baker, and J.~A. Carrillo.
\newblock A local continuum model of cell-cell adhesion.
\newblock {\em arXiv preprint arXiv:2206.14461}, 2022.
\newblock To appear in SIAM Journal on Applied Mathematics.

\bibitem{smith1973cells}
J.~Smith and L.~Martin.
\newblock Do cells cycle?
\newblock {\em Proceedings of the National Academy of Sciences},
  70(4):1263--1267, 1973.

\bibitem{weber2014quantifying}
T.~S. Weber, I.~Jaehnert, C.~Schichor, M.~Or-Guil, and J.~Carneiro.
\newblock Quantifying the length and variance of the eukaryotic cell cycle
  phases by a stochastic model and dual nucleoside pulse labelling.
\newblock {\em PLoS Computational Biology}, 10(7):e1003616, 2014.

\bibitem{yates2017multi}
C.~A. Yates, M.~J. Ford, and R.~L. Mort.
\newblock {A multi-stage representation of cell proliferation as a Markov
  process}.
\newblock {\em Bulletin of Mathematical Biology}, 79:2905--2928, 2017.

\bibitem{kynaston2022equivalence}
J.~C. Kynaston, C.~Guiver, and C.~A. Yates.
\newblock Equivalence framework for an age-structured multistage representation
  of the cell cycle.
\newblock {\em Physical Review E}, 105(6):064411, 2022.

\bibitem{liu2023parameter}
Y.~Liu, K.~Suh, P.~K. Maini, D.~J. Cohen, and R.~E. Baker.
\newblock Parameter identifiability and model selection for partial
  differential equation models of cell invasion.
\newblock {\em Journal of the Royal Society Interface}, 21(212):20230607, 2024.

\end{thebibliography}
\end{document}